\documentclass[paper=A4,12pt,abstract=true,numbers=noenddot]{scrartcl}
\pdfoutput=1

\makeatletter
\DeclareOldFontCommand{\rm}{\normalfont\rmfamily}{\mathrm}
\DeclareOldFontCommand{\sf}{\normalfont\sffamily}{\mathsf}
\DeclareOldFontCommand{\tt}{\normalfont\ttfamily}{\mathtt}
\DeclareOldFontCommand{\bf}{\normalfont\bfseries}{\mathbf}
\DeclareOldFontCommand{\it}{\normalfont\itshape}{\mathit}
\DeclareOldFontCommand{\sl}{\normalfont\slshape}{\@nomath\sl}

\usepackage[usenames,dvipsnames]{color}
  \definecolor{hgreen}{rgb}{0,.3,0}
  \definecolor{hred}{rgb}{.3,0,0}
  \definecolor{hblue}{rgb}{0,0,.3}
  \definecolor{LightGray}{gray}{0.95}
\usepackage[
	    colorlinks=true,
	    linkcolor=hblue,
	    citecolor=hgreen,
	    filecolor=hblue,
	    urlcolor=hred
	    ]{hyperref}
\usepackage[numbers,sort&compress]{natbib}
\usepackage{pdfpages}
\usepackage{graphicx}
\usepackage{mathrsfs}
\usepackage{amsmath}
\usepackage{array}
\usepackage{amssymb}
\usepackage{braket}
\usepackage{bbold}
\usepackage{slashed}
\usepackage{cleveref}
\usepackage{subfig}
\usepackage[titletoc,title]{appendix}
\usepackage[affil-it ]{authblk}
\usepackage{soul}

\newcommand{\beq}{\begin{equation}}
\newcommand{\eeq}{\end{equation}}

\setcapindent{1em}
\setkomafont{captionlabel}{\bf}
\setkomafont{caption}{\it}
\numberwithin{equation}{section}

\usepackage{fancyhdr}
\fancypagestyle{preprint}{
  \fancyhead[LE,RO]{CERN-TH-2017-240\\FERMILAB-PUB-17-523-T\\MITP/17-083\\ZU-TH-35/17}

}

\begin{document}
\renewcommand\Authands{, }
\unitlength = 1mm

\title{\LARGE Light Resonances and the Low-$\boldsymbol{q^2}$ Bin of $\boldsymbol{R_{K^*}}$}
\date{\normalsize \today}

\author[a]{\large Wolfgang~Altmannshofer%
        }

\author[b]{\large Michael~J.~Baker%
        }

\author[a]{\large Stefania~Gori%
        }

\author[c]{\large Roni~Harnik%
        }

\author[d,e,g]{\large Maxim~Pospelov%
        }

\author[f]{\large Emmanuel~Stamou%
        }

\author[g]{\large Andrea~Thamm%
        }

\affil[a]{\normalsize Department of Physics, University of Cincinnati, Cincinnati, OH~45221, USA}
\affil[b]{\normalsize Physik-Institut, Universit\"at Z\"urich, 8057 Z\"urich, Switzerland}
\affil[c]{\normalsize Theoretical Physics Department, Fermilab, Batavia, IL~60510, USA}
\affil[d]{\normalsize Department of Physics and Astronomy, University of Victoria, Victoria, BC~V8P~5C2, Canada}
\affil[e]{\normalsize Perimeter Institute for Theoretical Physics, Waterloo, ON~N2L~2Y5, Canada}
\affil[f]{\normalsize Enrico Fermi Institute, University of Chicago, Chicago, IL~60637, USA}
\affil[g]{\normalsize Theoretical Physics Department, CERN, 1211 Geneva, Switzerland}

\maketitle
\thispagestyle{preprint}



\begin{abstract}
LHCb has reported hints of lepton-flavor universality violation in the
rare decays $B \to K^{(*)} \ell^+\ell^-$, both in high- and low-$q^2$ bins.
Although the high-$q^2$ hint may be explained by new short-ranged
interactions, the low-$q^2$ one cannot.
We thus explore the possibility that the latter is explained by a new
light resonance.
We find that LHCb's central value of $R_{K^*}$ in the low-$q^2$ bin
is achievable in a restricted parameter space of new-physics scenarios in which
the new, light resonance decays preferentially to electrons
and has a mass within approximately $10$\,MeV of the di-muon threshold.
Interestingly, such an explanation can have a kinematic origin and
does not require a source of lepton-flavor universality violation.
A model-independent prediction is a narrow peak in the differential
$B \to K^* e^+e^-$ rate close to the di-muon threshold.
If such a peak is observed, other observables, such as the differential
$B \to K e^+e^-$ rate and $R_K$, may be employed
to distinguish between models.
However, if a low-mass resonance is not observed and the low-$q^2$
anomaly increases in significance, then the case for an experimental
origin of the lepton-flavor universality violating anomalies would
be strengthened.
To further explore this, we also point out that, in analogy to $J/\psi$ decays,
$e^+e^-$ and $\mu^+\mu^-$ decays of $\phi$ mesons can be used
as a cross check of lepton-flavor universality
by LHCb with $5$\,fb$^{-1}$ of integrated luminosity.
\end{abstract}



\section{Introduction}
The gauge sector of the Standard Model (SM) exhibits exact flavor
universality, which is only broken by the Yukawa couplings of the
quarks and leptons with the Higgs boson. One of the best ways to test
this property of the SM is to measure semi-leptonic neutral current
decays of $B$ mesons. In the SM, these decays are induced at
one-loop level and are additionally suppressed by the
Glashow--Iliopoulos--Maiani (GIM) mechanism.
For these decays, observables that are sensitive to lepton-flavor universality (LFU)
are ratios of decay rates to muons and electrons, i.e.,
\beq
R_M=\frac{{\rm{BR}}(B\to M \mu^+\mu^-)}{{\text{BR}}(B\to M e^+e^-)},~~~M=K,K^*,X_s,...
\eeq
Recently, the LHCb collaboration determined~\cite{Aaij:2014ora,Aaij:2017vbb}
\begin{align}\label{eq:RKexp}
 R_K &\equiv \frac{\text{BR}(B \to K \mu^+\mu^-)}{\text{BR}(B \to K e^+e^-)} = 
 0.745 ^{+0.090}_{-0.074} \pm 0.036\,,\,~ \text{for} ~ q^2 \in [1,6]~\text{GeV}^2 ~, \\\label{eq:RKstarexp}
 R_{K^*} &\equiv \frac{\text{BR}(B \to K^* \mu^+\mu^-)}{\text{BR}(B \to K^* e^+e^-)} =
 \begin{cases}
 0.66 ^{+0.11}_{-0.07} \pm 0.03\,,~~ \text{for} ~ q^2 \in [0.045,1.1]~\text{GeV}^2 ~, \\
 0.69 ^{+0.11}_{-0.07} \pm 0.05\,,~~  \text{for} ~ q^2 \in [1.1,6]~\text{GeV}^2 ~,
\end{cases}
\end{align}
where $q^2$ is the di-lepton invariant mass squared.
The SM predictions for these observables have small, percent-level uncertainties.
Away from the di-muon threshold, $q^2 = 4m_\mu^2 \simeq 0.045$~GeV$^2$,
$R_K^\text{SM}$ and $R_{K^*}^\text{SM}$ are $1$ with high
precision~\cite{Hiller:2003js,Bordone:2016gaq}.
$R_{K^*}^\text{SM}$ in the low-$q^2$ bin is slightly
below $1$, mainly due to phase space effects~\cite{Bordone:2016gaq}:
\begin{align}
 R_K^\text{SM} & = 1.00 \pm 0.01\,,\quad~ ~\, \text{for} ~ q^2 \in [1,6]~\text{GeV}^2 ~, \\
 R_{K^*}^\text{SM} &=
 \begin{cases}
 0.91 \pm 0.03\,,\quad  \text{for} ~ q^2 \in [0.045,1.1]~\text{GeV}^2 ~, \\
 1.00 \pm 0.01\,,\quad  \text{for} ~ q^2 \in [1.1,6]~\text{GeV}^2 ~.
\end{cases}
\end{align}
These predictions are in some tension with the LHCb measurements in eqs.~\eqref{eq:RKexp} and \eqref{eq:RKstarexp}.
Combining the errors in quadrature, one finds an $\sim 2.6\sigma$
tension in $R_K$, and an $\sim 2.4\sigma$ and $\sim 2.5\sigma$
tension in the two bins for $R_{K^*}$.

If the discrepancies between measurements and SM predictions are
due to New Physics (NP) from four-fermion contact interactions,
the ratio $R_{K^*}$ is expected to have a non-trivial $q^2$ dependence.
At low di-lepton invariant mass, the $B \to K^* \ell^+ \ell^-$ rates
are dominated by a $1/q^2$ enhanced photon contribution, which strongly
dilutes NP effects in the low-$q^2$ bin.
Model independent
analyses~\cite{Altmannshofer:2017yso,Capdevila:2017bsm,DAmico:2017mtc,Geng:2017svp,Ciuchini:2017mik}
find that a NP contact interaction that explains $R_K$
and $R_{K^*}$ in the high-$q^2$ bins affects $R_K^{*}$ in the
low-$q^2$ bin typically by at most $10\%$.
We are, therefore, led to explore the possibility that the low-$q^2$
discrepancy in $R_{K^*}$ may be a hint for new light degrees of
freedom, which cannot be described by an effective Lagrangian with
only SM fields (see, however, also ref.~\cite{Bardhan:2017xcc}).

The possible effects of resonances below the electroweak scale
on LFU in $B \to K^{(*)} \ell^+ \ell^-$ have been previously
considered in refs.~\cite{Fuyuto:2015gmk,Datta:2017pfz,Sala:2017ihs,Ghosh:2017ber,Alok:2017sui,Bishara:2017pje,Datta:2017ezo,Babu:2017olk}.
In this work, we point out that a light, new resonance can affect the
low-$q^2$ bin of $R_{K^*}$ only in a very restricted range of
parameter space once all relevant constraints are
taken into account.
If the resonance has a mass significantly below the di-muon threshold,
it affects $R_{K^*}$ from an off-shell exchange.
We find, however, that the related two-body decays of $B$ mesons into
final states containing the resonance on-shell typically
oversaturate the total $B$ width. We thus exclude such a scenario.
If the resonance mass is close to or above the di-muon threshold,
strong constraints exist from the existing
measurements of the differential $B \to K^* e^+e^-$ rate~\cite{Aaij:2015dea} and from di-muon resonance searches in the
$B \to K^* \mu^+\mu^-$ decay~\cite{Aaij:2015tna}.

Our main result is that a light new resonance can produce a
suppression of $R_{K^*}$ in the low-$q^2$ bin only if the resonance
decays preferentially to electrons and its mass is within
approximately $10$~MeV of the di-muon threshold.
Such a situation can occur either because the resonance couples
non-universally to charged leptons or because its decay to muons
is kinematically forbidden even if its coupling is universal, e.g.,
dark-photon models.
This leads to testable consequences for other LHCb measurements.
In particular, it implies that the differential $B \to K^* e^+e^-$
rate close to the di-muon threshold features a peak that should
be searched for experimentally.
Analogously, the $B_s \to \phi e^+e^-$ spectrum has to feature
a peak close to the di-muon threshold of the same relative size.
A peak should also be present in the differential
$B \to K e^+e^-$ rate close to the di-muon threshold.
While $K^*$ and $\phi$ are vectors, $K$ is a pseudoscalar.
Therefore, the size of the peak in $B \to K e^+e^-$ is
model dependent and allows us to distinguish between different flavor
violating interactions of the resonance to bottom and strange quarks.

The connection between the deviation in the low-$q^2$ bin of $R_{K^*}$ and
the peaks in the $B \to K^* e^+e^-$ and $B_s \to \phi e^+e^-$ spectra
is robust.
This allows us to further conclude that if the low-$q^2$ deviation
persists and becomes statistically significant, but no peak is observed,
the case for a systematic experimental origin of the deviation would
be strengthened.
This will have implications for the interpretation
of any anomaly in the high-$q^2$ bin, if it persists.

The paper is organized as follows:
In \cref{sec:model_independent}, we show how a light, new
resonance can affect the low-$q^2$ bin of $R_{K^*}$ taking into
account all relevant experimental constraints.
This analysis is model-independent since it does not
depend on how the resonance couples to bottom and strange quarks.
We consider both off-shell and on-shell explanations and argue that
only the latter is consistent with other observations.
We also discuss the model-independent implications for the
$B\to K^*e^+e^-$ and $B_s\to \phi e^+e^-$ spectra.
In \cref{sec:model_dependent}, we consider several models,
focusing on new vector resonances just below the di-muon threshold.
We analyse different ways to couple the resonance to the flavor
changing quark current and show the corresponding model dependent
implications for the $B \to K e^+e^-$ decay.
In \cref{sec:newLFU}, we propose additional LFU measurements
for the LHCb experiment that could lead to further insights into the
origin of the low-$q^2$ anomaly.
Finally, in \cref{sec:conclusions} we discuss and summarize
our results.
In \cref{sec:3body} we elaborate on the off-shell case,
and in \cref{sec:2bodyWidths} we report the form factors used in our analysis.


\section{Model-Independent Analysis\label{sec:model_independent}}
In this section, we discuss the impact of a light, new resonance,
$X$, in $R_{K^*}$, keeping the discussion as model
independent as possible.

\subsection{Off-shell effect of a light resonance\label{sec:off-shell}}

The off-shell exchange of a resonance far below the di-muon threshold
can in principle contribute
to the $B \to K^* \ell^+ \ell^-$ rate in the low-$q^2$ bin.
The propagator is approximately proportional to $1/q^2$, which
enhances the off-shell contribution at low $q^2$ (like the SM photon).
We thus expect such off-shell exchanges to have a high impact on
measurements at low $q^2$, which could account for the anomaly
in the low-$q^2$ bin of $R_{K^*}$.
However, we show here that such a setup is unlikely to
satisfy existing experimental constraints.

To illustrate this point, we consider a very light resonance, $X$,
with a mass far below the low-$q^2$ bin of $R_{K^*}$, i.e., $m_X^2\ll 0.045$\,GeV$^2$,
that couples to leptons (with coupling $g_\ell$, $\ell = \mu, e$)
and off-diagonally to bottom and strange quarks.
If the off-shell exchange of $X$ produces
a visible effect in $R_{K^*}$, then this would typically
imply a two-body inclusive $B \to X_s X$ width
that exceeds the total $B$ width.
For example, if we assume that $X$ has a flavor changing
dipole interaction\footnote{%
The qualitative conclusions remain the same for different choices
of the particle $X$ and its interactions with fermions.},
we estimate that
\begin{equation}\label{eq:BRbsrho}
\frac{ \Gamma(B \to X_s X)}{\Gamma_{B,\text{tot}}^\text{SM}}
\sim \frac{e^2}{4 g_\ell^2} (\Delta R_{K^*})^2 \times \text{BR}(B \to X_s \gamma) \simeq 800\% \times
\left( \frac{0.3 \cdot 10^{-3}}{g_\ell} \right)^2 \left( \frac{\Delta R_{K^*}}{0.3} \right)^2\,,
\end{equation}
where $\Gamma_{B,\text{tot}}^\text{SM}$ is the total width of the $B$
meson in the SM, $\Delta R_{K^*}\equiv R_{K^*}^\text{SM} - R_{K^*}$
(in the low-$q^2$ bin), and where we have used
BR$(B \to X_s \gamma) = (3.32 \pm 0.15)\cdot 10^{-4}$~\cite{Amhis:2016xyh}.
Given that the coupling of light ($\sim 10$'s of MeV) new degrees of
freedom to electrons and muons are constrained to be
$\lesssim 10^{-3}$ (see \cref{fig:lepton_couplings} in
\cref{sec:3body}), the $B \to X_s X$ decay width
typically exceeds the
experimentally determined total $B$ width by a factor of a few, which
excludes such a scenario.
For the derivation of \cref{eq:BRbsrho} we assumed that the resonance
couples only to one type of lepton.
Barring cancellations, the same argument leads to even more stringent
constraints if we assume couplings to both muons and electrons.
We quantify our argument in detail for a vector resonance in
\cref{sec:3body}.

\subsection{On-shell production of a light resonance} \label{sec:on-shell}

Having argued that the off-shell exchange of a light resonance cannot
affect the low-$q^2$ bin of $R_{K^*}$ in an appreciable way, we now
discuss scenarios in which on-shell production of the resonance
($B \to K^* X$ with $X \to \ell^+ \ell^-$) affects the low-$q^2$
bin.
In the case of a narrow resonance,
this is possible as long as the mass of the resonance is inside
the $[0.045,~1.1]$~GeV$^2$ bin, up to experimental resolution effects.
In the on-shell approximation there is no interference with the SM
$b \to s \ell \ell$ amplitudes, so the resonance can only enhance
the $B \to K^* \ell^+ \ell^-$ rates.
Therefore, in order to explain $R_{K^*}$ in this scenario, the resonance
has to decay more often into electrons than into muons, i.e.,
BR$(X \to e^+e^-) >$ BR$(X \to \mu^+\mu^-)$.

In general, the scenario can be model independently defined by the following
set of parameters:
(i) the mass of the resonance, $m_X$;
(ii) the $B$ meson branching ratio BR$(B \to K^* X)$;
(iii) the leptonic branching ratios of the resonance, BR$(X \to e^+e^-)$ and
BR$(X \to \mu^+\mu^-)$;
(iv) the total width of the resonance $\Gamma^X_\text{tot}$.

We will find that the mass of $X$ has to be close to the di-muon threshold.
Far below the threshold, the effect in $R_{K^*}$ becomes negligible, while far
above the threshold the constraints from the measured
$B \to K^* e^+e^-$ spectrum and searches for $B \to K^* X(\to \mu^+\mu^-)$
are severe.
In the following, we therefore focus on the case of $X$ masses for which
the decay to $\tau$'s or to two or more hadrons is kinematically
forbidden.
The total $X$ width is then the sum of the partial width into the visible
final states of electrons and muons, as well as the width into invisible
final states like neutrinos and any other kinematically accessible decay
channel of the $X$ to ``dark'', non-SM particles.\footnote{%
We do not consider the decay $X\to \pi_0 \gamma$, which
has a tiny branching ratio in typical models.  We also do not consider the decay 
$X\to \gamma \gamma$ that is possible if $X$ is a (pseudo)scalar.}
We work in the limit of narrow width, $\Gamma_\text{tot}^X \ll m_X$.
The width of $X$ is bounded from below, as the leptons and
the $K^*$ are observed to originate from the same vertex~\cite{Aaij:2017vbb}.
Demanding that the $X$ decays promptly ($c\tau \lesssim 2$mm)
and using a typical boost factor of 200,\footnote{We estimate the boost factor
using a mean energy of $80$~GeV for the $B$ mesons at LHCb~\cite{Altarelli:2008xy}.}
we find $\Gamma^X_\text{tot} \gtrsim 0.02$ eV.
This is compatible with the narrow-width assumption for the range of
masses we consider.

The new resonance then affects the $B \to K^* \ell^+ \ell^-$
branching ratios in a given bin of di-lepton
invariant mass $[q_\text{min}^2,~q_\text{max}^2]$ in the following way
\begin{equation}
\langle \text{BR}_{\ell\ell}\rangle\bigr\vert^{q_{\rm max}}_{q_{\rm min}}
=
\langle \text{BR}^{\rm SM}_{\ell\ell}\rangle\bigr\vert^{q_{\rm max}}_{q_{\rm min}} +
\text{BR}(B\to K^*X)\cdot
	\text{BR}(X\to\ell^+\ell^-)\cdot
	{\cal G}^{(r_\ell)}(q_{\rm min}, q_{\rm max} ) \,.
\label{eq:diffBRll}
\end{equation}
The function ${\cal G}^{(r_\ell)}(q_{\rm min}, q_{\rm max} )$ models
the imperfect di-lepton mass resolution of the LHCb detector. We assume a
Gaussian smearing such that
\begin{equation}
{\cal G}^{(r_\ell)}(q_{\rm min}, q_{\rm max})  = \frac{1}{\sqrt{2\pi} r_\ell }
\int_{q_{\rm min}}^{q_{\rm max}} d|q| e^{-\frac{(|q|-m_X)^2}{2 r_\ell^2}}\,.
\end{equation}
For the resolutions we use $r_e = 10$\,MeV for electrons~\cite{Ilten:2015hya}
and $r_\mu =2$\,MeV for muons~\cite{Aaij:2014jba}. We neglect the
dependence of the mass resolution on $q^2$, as we always consider a very
narrow range of masses for $X$.

The NP prediction for $R_{K^*}$ in the bin
$[q_\text{min}^2,q_\text{max}^2]$ is then determined by the
corresponding modified branching ratios
\begin{equation}
R_{K^*} = \langle \text{BR}_{\mu\mu}\rangle\bigr\vert^{q_{\rm max}}_{q_{\rm min}} \Big/
\langle \text{BR}_{ee}\rangle\bigr\vert^{q_{\rm max}}_{q_{\rm min}} ~.
\end{equation}
We use {\tt flavio}~\cite{flavio} to compute the SM predictions and
uncertainties
of $R_{K^*}$ and the branching ratios
$\langle \text{BR}^{\rm SM}_{\ell\ell}\rangle\bigr\vert^{q_{\rm max}}_{q_{\rm min}}$.

As long as the mass of the new resonance is not more than
$\mathcal O(r_e) = \mathcal O(10~\text{MeV})$ outside the
lower edge of the $[0.045,1.1]$~GeV$^2$ bin,
the $B \to K^* X$ and $X \to \ell^+\ell^-$
branching ratios can be adjusted to account for the $R_{K^*}$ value
measured by LHCb.
Various other measurements constrain the NP parameter space.
The most stringent constraints are:
\begin{itemize}
\item The LHCb search for a resonance in the di-muon invariant mass
spectrum in the $B\to K^*X (\to \mu^+\mu^-)$ decay~\cite{Aaij:2015tna}. This
search places very stringent upper limits on the product
BR$(B \to K^*X) \times$ BR$(X \to \mu^+\mu^-)$, 
which are given as a function of
the $X$ mass and the $X$ width.
If $X$ is to explain the low-$q^2$ bin of $R_{K^*}$,
the bounds for a promptly decaying $X$ apply.
No bound can be obtained from this search for $m_X < 2 m_\mu$,
where BR$(X \to \mu^+\mu^-) = 0$.
\item The differential branching ratio of $B\to K^* e^+e^-$ measured
by LHCb~\cite{Aaij:2015dea} constrains the product of
BR$(B \to K^*X) \times$ BR$(X \to e^+e^-)$ for resonance masses
below and above the di-muon threshold.
LHCb presents measurements of six bins of $q^2$, ranging from $0.0004$~GeV$^2$ to
$1$~GeV$^2$ \cite{Aaij:2015dea}.
Interestingly, a small excess of $B\to K^* e^+e^-$ events is observed
in the $q^2$ bin below the di-muon threshold, leading to a slight
preference for a non-zero BR$(B \to K^*X) \times$ BR$(X \to e^+e^-)$
for $m_X < 2m_\mu$.
\item The bounds on BR$(B\to K^*\nu\bar\nu)$ obtained at the $B$
factories~\cite{Lutz:2013ftz,Lees:2013kla} are relevant for the case
in which the resonance has a sizeable branching ratio into
invisible final states.
The most stringent bound is obtained by Belle~\cite{Lutz:2013ftz}; it reads
BR$(B \to K^*X) \times$ BR$(X \to \text{invisible}) < 5.5 \cdot 10^{-5}$ at $90\%$
Confidence Level (C.L.).
\end{itemize}

For the numerical analysis, we construct a $\chi^2$ function based on
a gaussian likelihood function that contains
the low-$q^2$ bin of $R_{K^*}$, the limits from the $B\to K^* X(\to\mu^+\mu^-)$ search,
the $B\to K^* e^+e^-$ distribution, and the $B\to K^*\nu\bar\nu$ bound.
To account for the asymmetric error of $R_{K^*}$ we use the positive (negative) side of
error if the $R_{K^*}$ prediction lies above (below) the experimental central value.
From ref.~\cite{Aaij:2015tna} we extract the bound on prompt $X$
decays BR$(B\to K^*X)\times$BR$(X\to\mu^+\mu^-)< 3\cdot 10^{-9}$ at $95\%$
C.L., which we implement in the $\chi^2$ for all masses $m_X>2m_\mu$ close to
the di-muon threshold.
We take into account all $q^2$ bins
measured in the LHCb analysis~\cite{Aaij:2015dea}
of $B\to K^{*}e^+e^-$.
The theory uncertainties in this $q^2$ region are mainly due to form
factors and CKM elements. We, therefore, assume that these uncertainties
are $90\%$ correlated across the bins.
Efficiency effects are estimated by comparing the
SM prediction from {\tt flavio} with the ones presented in the LHCb
analysis~\cite{Aaij:2015dea}.
To capture possible uncertainties of this procedure,
we inflate the theory uncertainties from {\tt flavio} by a factor of
$1.5$ to be conservative.
Taking  into account the correlation, we add the theory errors in
quadrature with the experimental errors. We have checked that
choosing a different level of correlation in the theory uncertainties
does not lead to qualitative changes in our results.

For a given set of NP parameters, we plot contours of
$\Delta\chi^2=\chi^2-\chi^2_{\rm min}$ corresponding to the preferred regions at
$68.27\%$ and $95.45\%$ C.L., and give both $\chi^2_{\rm min}$ and
$\chi^2_{\text{SM}}$ for comparison.
We also show separately the preferred $68.27\%$ C.L.\ region for $R_{K^*}$
in the low-$q^2$ bin ignoring all constraints.
For the constraints from $B\to K^{*} X (\to \mu^+\mu^-)$, $B\to K^{*}e^+e^-$,
and $B\to K^{*}\nu\bar\nu$ we shade the part of the parameter
space that is excluded at $95\%$ C.L.
We find that once the above mentioned constraints are taken into account,
the discrepancy in the low-$q^2$ bin of $R_{K^*}$ can only be addressed in
a very restricted range of NP parameter space.
We first illustrate this in a simple benchmark scenario, in which we
identify the resonance with a dark photon, i.e., $X\Rightarrow A'$.
We then discuss the viable parameter space in the case of a generic resonance.

\subsubsection{Dark photon -- LFU violation without LFU violation}

\begin{figure}[t]
\includegraphics{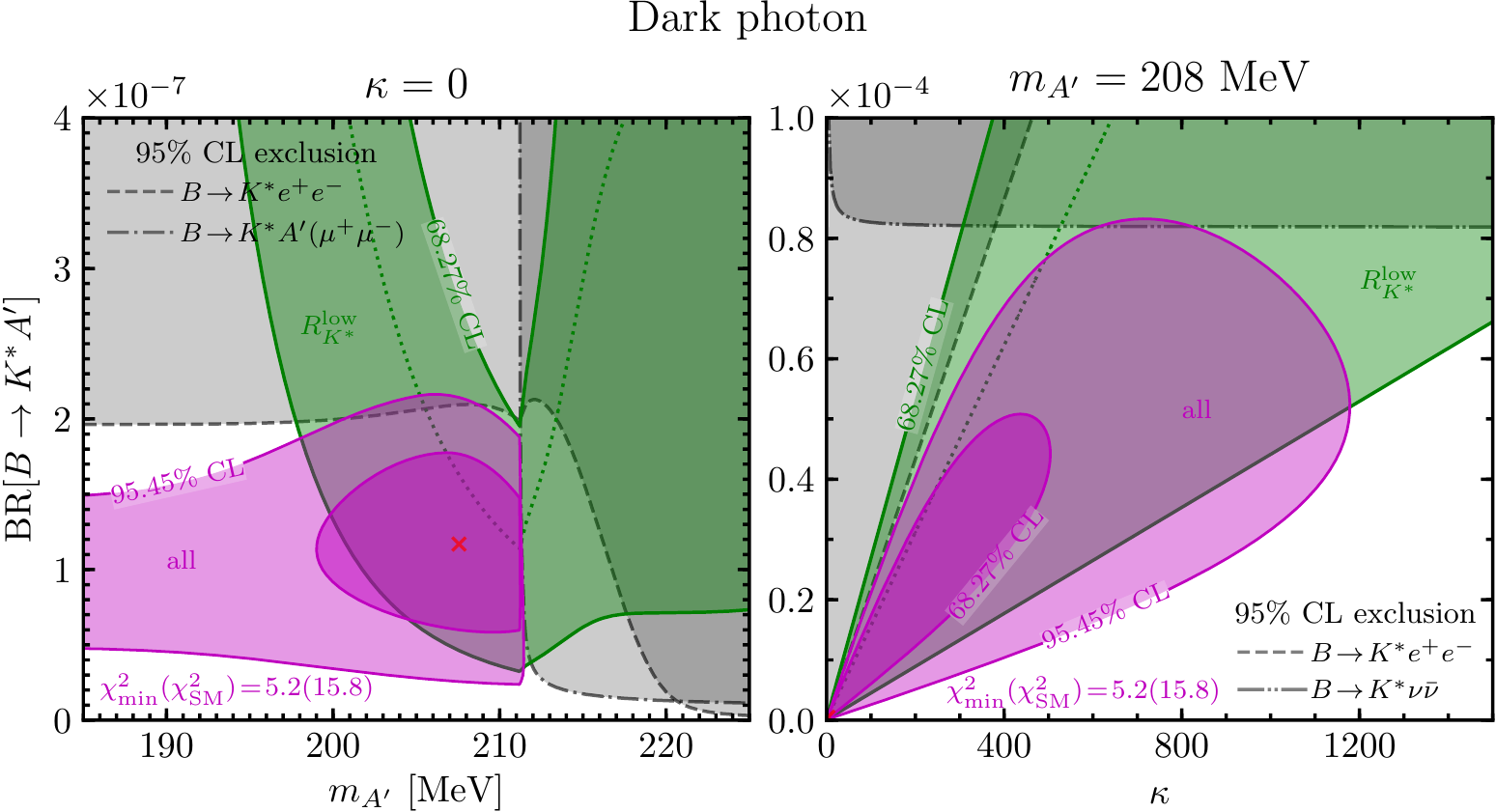}
\caption{Preferred regions of parameter space for a dark-photon
explanation of the low-$q^2$ bin of
$R_{K^*}$. In the left plot, the dark photon is assumed to decay
$100\%$ to electrons and muons; the dark-photon mass and BR($B \to K^* A'$)
are varied. In the right plot, the dark-photon mass is fixed
to $m_{A'} = 208$~MeV; the BR($B \to K^* A'$) and the invisible
width (parameterized by $\kappa$, see text) are varied.
The red cross at BR($B \to K^* A') = 1.2\cdot 10^{-7}$ and
$m_{A'} = 208\,$MeV (left), and BR($B \to K^* A') = 1.2 \cdot 10^{-7}$ and
$\kappa = 0$ (right) are the best-fit values in each case.
\label{fig:darkphoton}}
\end{figure}

If the resonance is a dark photon, $A'$, its branching ratios to
electrons and muons are fixed by the dark-photon mass, $m_{A'}$,
its total width, $\Gamma_{\rm tot}^{A'}$,
and either the kinetic-mixing parameter $\epsilon$ or equivalently
the dark-photon partial width to non-SM particles, $\Gamma_{\rm other}^{A'}$.
In the mass range we consider, the total width is given by
\begin{equation}
\Gamma^{A'}_{\text{tot}} =  \Gamma^{A'}_{ee} + \Gamma^{A'}_{\mu\mu} + \Gamma^{A'}_{\text{other}}\,,
\end{equation}
with
\begin{equation}
\Gamma^{A'}_{\ell\ell} =
\epsilon^2 \frac{e^2}{12\pi} m_{A'}\left( 1+
2\frac{m_\ell^2}{ m_{A'}^2}\right)\sqrt{1-4\frac{m^2_\ell}{ {m_{A'}^2}}}
 \theta(m_{A'}^2-4m_\ell^2)\,.
\end{equation}
We find it convenient to parameterize
$\Gamma^{A'}_{\text{other}} = \kappa(\Gamma^{A'}_{ee} + \Gamma^{A'}_{\mu\mu})$.
In this parametrization, the dark-photon branching ratios to electrons and muons
are independent of $\epsilon$.
A dark-photon benchmark is then fully specified by
choosing\footnote{The kinetic mixing parameter $\epsilon$ determines
the total width of the dark photon. As long as $\epsilon$ is large
enough such that the dark photon decays promptly, the exact value of
$\epsilon$ is not relevant for our discussion.}
\begin{equation}
m_{A'},\quad \text{BR}(B\to K^{*}A'),\quad \kappa \,.
\end{equation}

In the left panel of \cref{fig:darkphoton} we consider the case
of $\kappa=0$ and show the constraints and preferred region in the
parameter space of $m_{A'}$ and BR$(B\to K^{*}A')$.
In green we show the preferred $68.27\%$ C.L.\,region for the low-$R_{K^*}$ bin
and in magenta the preferred 68.27\% and 95.45\% C.L.\,regions of the combined
$\chi^2$.
LHCb's constraints on $B\to K^{*}e^+e^-$ and $B\to K^{*}X(\to \mu^+\mu^-)$
exclude the shaded grey regions at $95\%$ C.L.
There is no constraint from $B\to K^{*}\nu\bar\nu$ here since $\kappa = 0$.
The best-fit point of the joint $\chi^2$ is at
$m_{A'} \simeq 208$\,MeV and BR$(B \to K^* A') \simeq 1.2 \cdot 10^{-7}$
(red cross in \cref{fig:darkphoton}).\footnote{
For comparison, note that in the SM the branching ratios in the low-$q^2$ bin are
$\langle \text{BR}^\text{SM}_{ee}\rangle|_{\text{low-}q^2} \simeq 1.3\cdot 10^{-7}$
and
$\langle \text{BR}^\text{SM}_{\mu\mu}\rangle|_{\text{low-}q^2} \simeq 1.2\cdot 10^{-7}$.
}
We see that the preferred region (magenta) is constrained
to be below and close to the di-muon threshold.
After profiling away the BR$(B\to K^*A')$ direction we find that
$m_{A'}\in[203, ~211]$ MeV at $68.27\%$ C.L.
The comparison of the minimum of the joint $\chi^2$, $\chi^2_{\text{min}}= 5.2$,
to the SM one, $\chi^2_\text{SM}= 15.8$, shows that
a dark photon in the preferred region describes low-$q^2$ data significantly
better than the SM alone. This is driven by an improved fit to $R_{K^*}$.

Next we turn on the partial width of $A'$ to light non-SM particles, i.e.,
$\kappa\neq 0$. The presence of these additional decay channels reduces the
branching ratios of $A'$ to electrons and muons.
Correspondingly, a larger BR$(B\to K^{*}A')$ is required to explain
the anomaly.
This is illustrated in the right panel of \cref{fig:darkphoton}
where we fix the dark-photon mass to $m_{A'} = 208$~MeV and show the
preferred region of parameter space in the $\kappa$ vs.\ BR$(B\to K^{*}A')$
plane.
We see that for large values of $\kappa > {\cal O}(10^3)$,
the constraint from $B \to K^* + \text{invisible}$ excludes an
explanation of $R_{K^*}$ by the dark photon. However, the point with $\kappa = 0$ is slightly preferred.

Interestingly, the dark-photon explanation of the low-$q^2$ bin
does not introduce any sources of LFU violation beyond the SM.
In this attractive, minimal scenario, the modification of $R_{K^*}$ arises
due to the difference of electron and muon mass. Note that the value of $R_{K^*}$  does not depend on the
kinetic-mixing parameter, $\epsilon$, as long as the dark photon decays promptly.
At a mass of $\sim$210~MeV the dark photon is constrained by the APEX, MAMI,
and BaBar experiments to have a mixing
$\epsilon\lesssim 10^{-3}$~\cite{Abrahamyan:2011gv,Essig:2013lka,Merkel:2014avp,Lees:2014xha,Alexander:2016aln}.
A dark photon with a coupling that saturates this limit has a decay length of about $80$ microns
including a typical Lorentz boost factor of 200~(see footnote 3).
This is fully compatible with the maximal displacement of $2$~mm seen in
the $R_{K^*}$ measurement~\cite{Aaij:2017vbb}.

\subsubsection{Generic resonance}

\begin{figure}[t]
\includegraphics{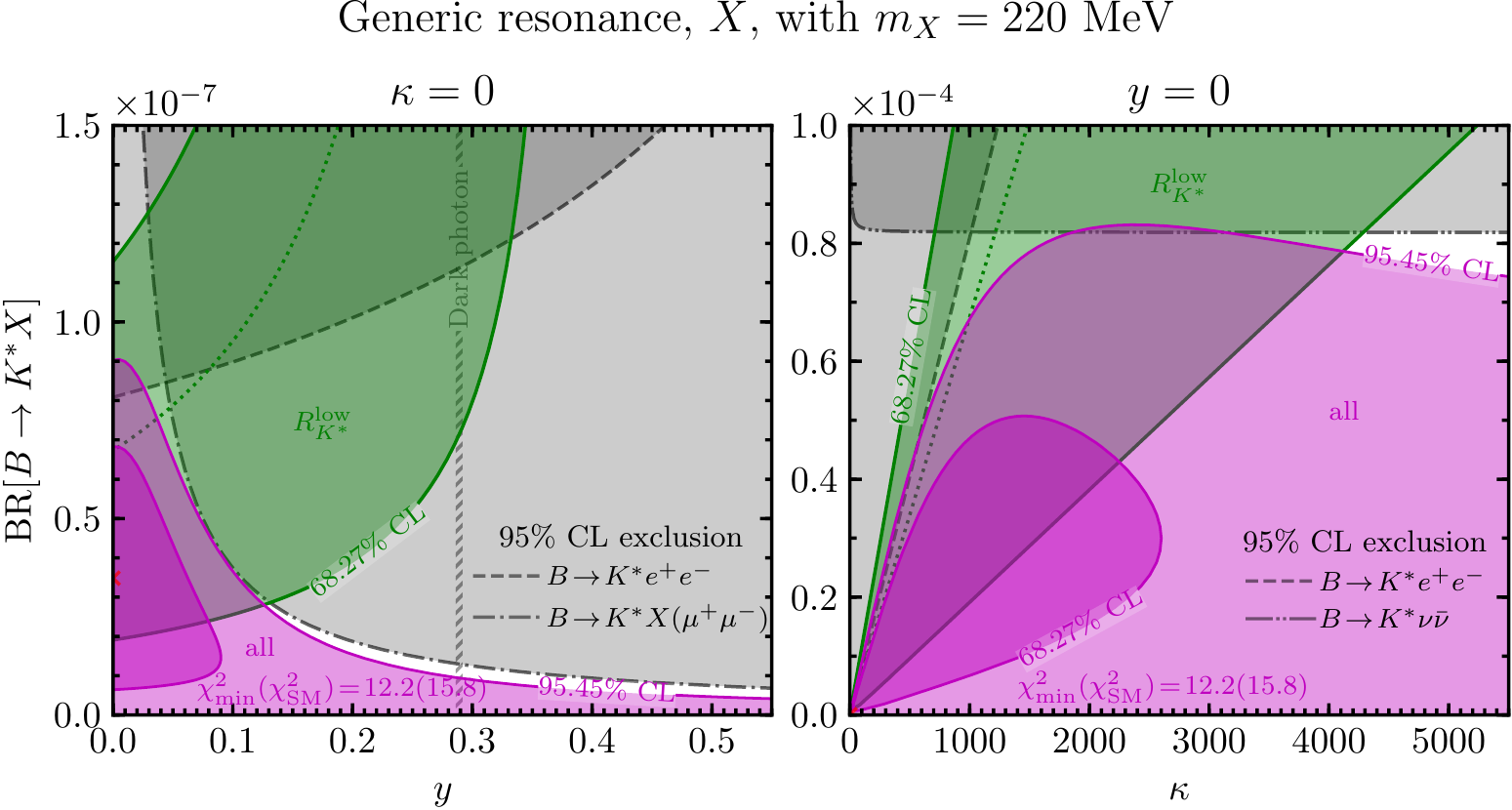}
\caption{Preferred regions of parameter space for a generic resonance
explanation of the low-$q^2$ bin of
$R_{K^*}$. The mass of the resonance is fixed to 220~MeV. In the
left plot the invisible branching ratio is set to zero ($\kappa=0$). In the right
plot the branching ratio to muons is set to zero ($y=0$).
The red crosses at BR$(B\to K^* X)=3.5 \cdot 10^{-8}$ and $y=0$ (left), and
BR$(B\to K^* X)=3.5 \cdot 10^{-8}$ and $\kappa=0$ (right) correspond to
the best-fit values.
\label{fig:zprimemassvariation}}
\end{figure}

In the generic case, we treat the electron and muon branching ratios of the
resonance as independent parameters.
Instead of introducing a Lagrangian, for which we would have to specify
the spin of the resonance and the chiral structure of its couplings, we
introduce the parameter $y\in[0,1]$ which interpolates between
the case of
BR$(X\to \mu^+\mu^-)=0$ for $y=0$ and BR$(X\to e^+e^-)=0$ for
$y=1$. We thus use the parameterization
\begin{align}
\text{BR}(X\to e^+e^-)      &= \frac{1}{1+\kappa}\cdot (1-y) \,,\\
\text{BR}(X\to \mu^+\mu^-)  &= \frac{1}{1+\kappa}\cdot y \,,\\
\text{BR}(X\to\text{other}) &= \frac{\kappa}{1+\kappa}\,.
\end{align}
The generic scenario is then fully specified by the parameter set
\begin{equation}
m_{X},\quad \text{BR}(B\to K^{*} X),\quad y,\quad \kappa \,.
\end{equation}

For a resonance mass below the di-muon threshold, i.e.,
$m_X < 2m_\mu$, the branching ratio to muons vanishes and,
thus, at these masses this scenario is identical to the dark-photon
model discussed in the previous section.
In \cref{fig:zprimemassvariation} we pick a
mass for the resonance above the di-muon threshold, $m_{X}= 220$ MeV.
In the left panel, we show the preferred region in
the space of $\text{BR}(B \to K^{*}X)$ and $y$, fixing $\kappa = 0$
corresponding to the case of no invisible decays.
We observe that a resonance with a larger branching ratio
to electrons than to muons, i.e., $y<0.5$, is preferred.
The dashed vertical line at $y = 0.29$
corresponds to the case of the dark-photon scenario discussed above.
In the right panel,
we vary $\text{BR}(B \to K^{*}X)$ and $\kappa$, fixing $y = 0$
corresponding to BR$(X\to \mu^+\mu^-)=0$.
As in the case of the dark photon, a large invisible branching ratio
is allowed.

We see that for $m_{X}= 220$ MeV, the minimum of the total
$\chi^2$ is significantly larger than for the dark-photon case above
($\chi^2_\text{min}=12.2$ and $5.2$, respectively) and corresponds to
BR$(B\to K^* X)=3.5 \cdot 10^{-8}$ and $y=0$ in the case of $\kappa=0$,
and to BR$(B\to K^* X)=3.5 \cdot 10^{-8}$ and $\kappa=0$ in the case of $y=0$
(red crosses in \cref{fig:zprimemassvariation}).
This is predominantly due to the tension between
the low-$q^2$ bin in $R_{K^*}$ and
the $B \to K^{*}e^+e^-$ constraint for this choice of $m_X$.
If we increase the $X$ mass to values above $220$~MeV,
the constraint from the $B \to K^{*}e^+e^-$ spectrum
becomes stronger excluding an explanation of the
low-$q^2$ anomaly in $R_{K^*}$.

\subsubsection{Model-independent predictions}

\begin{figure}[t]
\begin{center}
\includegraphics{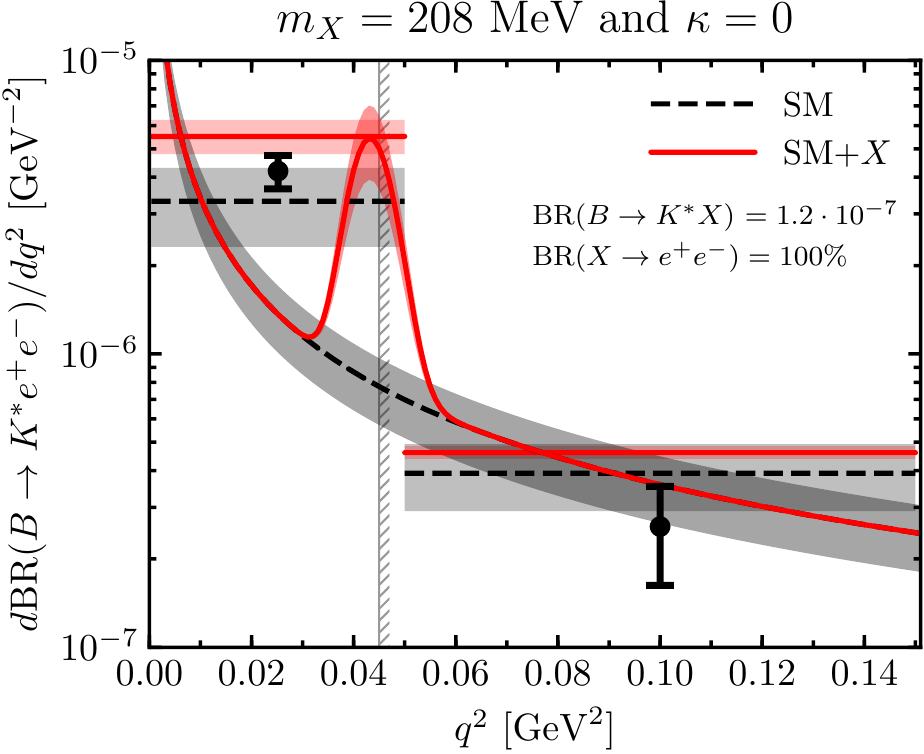}
\end{center}
\caption{The differential branching ratio $d$BR$(B\to K^* e^+ e^-)/dq^2$
in the SM (dashed black) and in the presence of a resonance with
the best-fit values for its mass and BR($B\to K^*X$) (solid red).
We also show the binned predictions together with
the LHCb measurements from ref.~\cite{Aaij:2015dea}.
The vertical line indicates the lower boundary of the low-$q^2$
bin in the $R_{K^*}$ measurement.
\label{fig:BRBKstaree}}
\end{figure}

As discussed above, any on-shell explanation of the low-$q^2$
bin of $R_{K^*}$ requires a resonance close to the di-muon threshold
decaying preferentially into electrons.\footnote{
In the past, a new particle in a very similar mass range had been proposed
in connection with flavor physics.
A light unspecified resonance was invoked as an explanation for
the anomalous clustering of events with di-muon mass at $214.3 \pm 0.5 $\,MeV
in the $\Sigma^+ \to p \mu^+\mu^-$ decay by the HyperCP
collaboration~\cite{Park:2005eka}.
Recent LHCb measurements of the same decay mode do not lend further
support to a hypothesis of a new $214$\,MeV particle \cite{LHCb:2016lop}.
A translation of these results to the $B$-meson case is not possible
in a model-independent way.}
A model-independent key prediction is therefore a peak in the
differential $B\to K^* e^+ e^-$ rate at a $q^2$ close to the
di-muon threshold.
For a resonance that decays
only to electrons $(y=0, \kappa = 0)$, the $68.27\%$ C.L.\ region for the mass is $m_X\in[203, 212]$~MeV.
If instead the resonance has a non-negligible decay mode into muons
(like the dark photon) the $68.27\%$ C.L.\ region is $m_X\in[203, 211]$~MeV.

For a resonance mass below the di-muon threshold, the size of
the peak is completely fixed. Above the di-muon threshold the
size of the peak scales as
$\text{BR}(X \to e^+e^-) / (\text{BR}(X \to e^+e^-) - \text{BR}(X \to \mu^+\mu^-))$.
In \cref{fig:BRBKstaree} we show the peak for the best
fit point below the di-muon threshold for $\kappa=0$.
We calculate the SM rate using \texttt{flavio}.
We see that the SM rate rises as $q^2 \rightarrow 0$, due to the contribution from
the photon pole.
We assume that the resonance is narrow and that
the spread in the NP events comes from the experimental 
resolution in electron reconstruction.  
Even taking this into account, the peak still rises prominently above
the background.
Also shown in the plot are SM and NP predictions of
BR$(B\to K^* e^+ e^-)$ for the $q^2$ bins measured by
LHCb~\cite{Aaij:2015dea} together with the experimental results.
More data and a finer $q^2$ binning should resolve the
peak if it is present.
An analogous peak with the same relative size is predicted in the
$B_s\rightarrow \phi e^+ e^-$ decay.


\section{Model-Dependent Implications\label{sec:model_dependent}}
We now consider possible operators that could induce the
$B\to K^* X$ transition for the case in which $X$ is a generic vector resonance, i.e.,
$X\Rightarrow V$.
In addition to constraining the Wilson coefficients, this
will allow us to make predictions for other observables, i.e.,
the differential rate of  $B\to K e^+e^-$ and $R_K$.
We shall find that a future precise measurement
of the differential $B \to K e^+e^-$ rate and of $R_K$ at low $q^2$
can distinguish the different operators if they are responsible
for the anomalous measurement of $R_{K^*}$ in the low-$q^2$ bin.

We concentrate here on vector resonances with masses just below the di-muon
threshold, such that the branching ratio into di-muons is zero.
We assume $100\%$ branching ratio into prompt electrons, neglecting
possible decays into a dark sector or neutrinos.
As we have shown in \cref{fig:darkphoton}, in the presence of a non-negligible invisible width of the resonance,
a larger $B\to K^* V$ branching ratio and, therefore, larger couplings
to quarks are required to compensate for the reduced $V \to e^+e^-$
branching ratio.

We consider flavor-violating couplings of the vector to bottom and
strange quarks up to dimension six
\begin{equation} \label{eq:Leff}
\mathcal{L}_{\rm eff} = \sum_{d=4,5,6} \left( \frac{C_{(d)}}{\Lambda^{d-4}} Q_{(d)} + \frac{C_{(d)}^\prime}{\Lambda^{d-4}} Q_{(d)}^\prime \right) ~+~ \text{h.c.}\,,
\end{equation}
where the operators are given by
\begin{align}
Q_{(4)}         &= (\bar s_L \gamma_{\mu} b_L) V^{\mu} \,,&
Q_{(4)}^\prime  &= (\bar s_R \gamma_{\mu} b_R) V^{\mu} \,,& \\
Q_{(5)}         &= (\bar s_L \sigma_{\mu\nu} b_R) V^{\mu\nu} \,,&
Q_{(5)}^\prime  &= (\bar s_R \sigma_{\mu\nu} b_L) V^{\mu\nu} \,,& \\
Q_{(6)}         &= (\bar s_L \gamma_{\mu} b_L) \partial_\nu V^{\mu\nu} \,,&
Q_{(6)}^\prime  &= (\bar s_R \gamma_{\mu} b_R) \partial_\nu V^{\mu\nu} \,, &
\end{align}
with $V_{\mu \nu} = \partial_\mu V_\nu - \partial_\nu V_\mu$.
In \cref{eq:Leff}, we also included the primed operators with a coupling of
opposite chirality with respect to the non-primed operators.
The widths and the analysis presented here for the primed operators are
equivalent to the ones of the corresponding non-primed operators.
We thus refrain from explicitly showing the results for the
primed operators.
In what follows, we shall assume that only one Wilson coefficient
contributes at a time. The presence of more than one operator
may produce additional interference effects.

Note that if one restricts to processes involving
on-shell $V$, even this minimal set of operators
is over-complete.
In particular, the free equation of motion for $V$ relates
$Q_{(6)} = m_V^2 Q_{(4)}$ and $Q'_{(6)} = m_V^2 Q'_{(4)}$.
Nevertheless, these operators are not fully equivalent as the amplitudes
with off-shell $V$ exchange differ for $Q_{(4)}$ and $Q_{(6)}$.
One can also wonder how a direct coupling of the vector to the
$bs$ current in $Q_{(4)}$ and $Q'_{(4)}$ is possible.
Recent studies have shown that these interactions do indeed arise
in models where a light $V$ couples to quark currents that are
not conserved when the SM mass terms and/or quantum anomaly effects
are taken into account~\cite{Dror:2017ehi,Dror:2017nsg}.
Models with direct flavour-universal couplings of $V$ to axial-vector
current of quarks tend to develop $Q_{(4)}$ at one loop, while models
with coupling of $V$ to any linear combinations of lepton and baryon
currents other than $B-L$ induce $Q_{(4)}$ at the two-loop level.

In concrete UV completions, the Wilson coefficients in \cref{eq:Leff}
will be suppressed by loop factors and couplings.
To make a connection to such UV models we pick a set of assumptions 
motivated by concrete examples and define a rescaled $\widetilde C_{(d)}$
for each $C_{(d)}$ as follows.

For $Q_{(4)}$ we assume that the interaction
is induced by the couplings of the vector to anomalous currents in
which case the coupling is two-loop suppressed \cite{Dror:2017ehi}
and we have
\begin{equation}
C_{(4)}= V_{tb} V^*_{ts} \left(\frac{e^2}{16 \pi^2}\right)^2
\widetilde{C}_{(4)}\,.
\label{eq:C4t}
\end{equation}
The concrete model of ref.~\cite{Dror:2017ehi} is gauged baryon number with a gauge
coupling $g_X$ and with a small kinetic mixing of the $U(1)_\mathrm{B}$ and the photon.
The translation from our coupling~$\widetilde C_{4}$ to this model is
$
\widetilde C_{(4)} = \frac{3}{\sin^4\theta_W} F\left(\frac{m_t^2}{m_W^2}\right) g_X \sim 10^2 g_X\,,
$
where $F(x)$ is a loop function of order one defined in~\cite{Dror:2017ehi}.
In other classes of models, this coupling can be induced at one-loop~\cite{Dror:2017nsg},
or at tree-level~\cite{Fox:2011qd,Altmannshofer:2014cfa}.

For $Q_{(5)}$ and $Q_{(6)}$ we assume that as in the SM the relevant couplings
are one-loop suppressed and that Minimal Flavor Violation aligns the flavor
structure of the couplings with the corresponding photonic
operators in the SM:
\begin{align}\label{eq:C5t}
\frac{C_{(5)}}{\Lambda} =&\, \frac{4 G_F}{\sqrt{2}} V_{tb} V^*_{ts} \frac{e^2}{16 \pi^2}\frac{m_b}{e}\widetilde{C}_{(5)} \frac{m_W^2}{\widetilde\Lambda^2} \,,\\\label{eq:C6t}
\frac{C_{(6)}}{\Lambda^2} =&\, \frac{4 G_F}{\sqrt{2}} V_{tb} V^*_{ts} \frac{e^2}{16 \pi^2} \frac{1}{e}\widetilde{C}_{(6)} \frac{m_W^2}{\widetilde\Lambda^2}\,.
\end{align}
Setting $\tilde\Lambda=m_W$ gives a Lagrangian in the normalization most
frequently employed for the effective Lagrangian in the SM.\footnote{%
To see this for $Q_{(6)}$, use the equation of motion for $V$ to relate
$Q_{(6)}$ to the semileptonic vector four-fermion operator appearing in the
SM ($O_9$).}

\subsection{Model interpretations of $\boldsymbol{B\to K^*}$ data}
The decay width $\Gamma(B \to K^* V)$ induced by each of
the operators in \cref{eq:Leff} is
\begin{align}
 \Gamma(B \to K^* V)\bigr\vert_{Q_{(4)}} &=  \frac{1}{64\pi} |C_{(4)}|^2 \frac{m_B^5 }{m_V^2 m_{K^*}^{2}} \left(1 - \frac{m_{K^*}^{2}}{m_B^2} \right)^{-2} \sqrt{\lambda}\, \mathcal{F}_{1}\left(\frac{m_{K^*}^{2}}{m_{B}^{2}},\frac{m_{V}^{2}}{m_{B}^{2}}\right) \, ,
	\label{eq:BKsrhoC4}\\
 \Gamma(B \to K^* V)\bigr\vert_{Q_{(5)}} &= \frac{1}{64\pi} \frac{|C_{(5)}|^2}{\Lambda^2}\frac{m_B^5}{m_{K^*}^{2}} \left(1 - \frac{m_{K^*}^{2}}{m_B^2} \right)^{-2} \sqrt{\lambda} \, \mathcal{F}_{2}\left(\frac{m_{K^*}^{2}}{m_{B}^{2}},\frac{m_{V}^{2}}{m_{B}^{2}}\right) \, ,
	\label{eq:BKsrhoC5}\\
 \Gamma(B \to K^* V)\bigr\vert_{Q_{(6)}} &= \frac{1}{64\pi} \frac{|C_{(6)}|^2}{\Lambda^4}\frac{m_B^5 m_V^2}{m_{K^*}^{2}}  \left(1 - \frac{m_{K^*}^{2}}{m_B^2} \right)^{-2} \sqrt{\lambda} \, \mathcal{F}_{1}\left(\frac{m_{K^*}^{2}}{m_{B}^{2}},\frac{m_{V}^{2}}{m_{B}^{2}}\right)\,,
	\label{eq:BKsrhoC6}
\end{align}
where we have defined the kinematical function
\begin{equation} \label{eq:kinfactor}
\lambda \equiv 1 + \frac{m_{K^*}^{4}}{m_{B}^{4}} + \frac{m_{V}^{4}}{m_{B}^{4}} - 2\frac{m_{K^*}^{2}}{m_{B}^{2}} - 2\frac{m_{V}^{2}}{m_{B}^{2}} - 2\frac{m_{K^*}^{2}}{m_{B}^{2}} \frac{m_{V}^{2}}{m_{B}^{2}} \, ,
\end{equation}
and used $B \to K^*$ form factors from ref.~\cite{Straub:2015ica}
to compute $\mathcal{F}_{1}$ and $\mathcal{F}_{2}$
(see eqs.~\eqref{eq:functionF1} and \eqref{eq:functionF2} in \cref{sec:2bodyWidths}).

Analogous expressions hold for the $C_{(d)}^\prime$ coefficients.
Notice that the same combination of form factors enter the decay
induced by $Q_{(4)}$ and $Q_{(6)}$.

\begin{table}
\renewcommand{\arraystretch}{1.5}
\centering
\begin{tabular*}{\textwidth}{@{\extracolsep{\fill}}lcccc}
\hline\hline
$\boldsymbol{d}$ 	&
$\boldsymbol{C_{(d)}^{\text{\bf best fit}}\bigr|_{\Lambda = 1\,\text{\bf TeV}}}$&
$\boldsymbol{\Lambda^{\text{\bf best fit}}\bigr|_{C_{(d)} = 1}}$&
$\boldsymbol{\widetilde{C}_{(d)}^{\text{\bf best fit}}\bigr|_{\widetilde{\Lambda} = 1\,\text{\bf TeV}}}$&
$\boldsymbol{\widetilde{\Lambda}^{\text{\bf best fit}}\bigr|_{\widetilde{C}_{(d)} = 1}}$\\[0.2em]
  \hline
  $4$ &  $\,\,1.6 \cdot 10^{-10}$	& --- 			 & $1.0 \cdot 10^{-2}$	& --- 	\\
  $5$ &  $2.4 \cdot 10^{-7}$		& $4.1 \cdot 10^{6}$ TeV & $3.3$ 		& $0.55$ TeV\\
  $6$ &  $3.7 \cdot 10^{-3}$ 		& $16$ TeV		 & $2.1 \cdot 10^{2}$	& $69$ GeV\\
   \hline\hline
\end{tabular*}
\caption{Values for Wilson coefficients and NP at the best-fit point.
For each operator the best-fit mass of the new resonance is approximately
208\,MeV. The normalization of the Wilson coefficients $\widetilde C_{(d)}$
is defined in eqs.~\eqref{eq:C4t}--\eqref{eq:C6t}.
\label{tab:bestfitpoint}}
\end{table}

We perform the $\chi^2$ fit
outlined in the previous section, including the constraints from
$B\to K^{*} V (\to \mu^+\mu^-)$, $B\to K^{*}e^+e^-$ and from the measured value
of $R_{K^*}$ in the low-$q^2$ bin.
In \cref{tab:bestfitpoint}, we list the best-fit value of the
Wilson coefficients, having fixed $\Lambda = 1\,\text{TeV}$, and
vice versa the value of $\Lambda$ having fixed the Wilson coefficient to be one.
This is shown both for $C_{(d)}$ and the rescaled $\widetilde C_{(d)}$.

We can now interpret the results of our best fit for the $\widetilde C_{(d)}$ in
\cref{tab:bestfitpoint} in the context of UV models, as well as in connection
with the high-$q^2$ bin of $R_{K^*}$.
\begin{itemize}
\item The dimension-four operators that can account for the low-$q^2$ anomaly
are well-behaved perturbative models, even if the coupling is suppressed
by two loops as in \cref{eq:C4t}.
It is notable that in the model of gauged baryon
number~\cite{Dror:2017nsg} this is achieved without any flavor violation neither
in the quark nor the lepton sector, the former being generated by the CKM matrix
and the latter by phase space.
\item The dimension-five, dipole-type operators can fit the low-$q^2$ deviation
for a NP at the TeV scale that is perturbative (coupling of order one)
and respects Minimal Flavor Violation.
Note that the same can be said for explanations of the high-$q^2$ deviation of $R_{K^*}$.
Turning this statement around, a generic TeV-scale explanation of the high-$q^2$ anomaly
can be augmented by a light mediator, for instance a dark photon, with a judiciously
chosen mass in order to explain the low-$q^2$ anomaly as well.
\item The dimension-six interpretation of the low-$q^2$ anomaly appears to be
disfavored with our UV asssumptions as a non-perturbative coupling
is required if the scale of NP is at the electroweak scale or higher.
\end{itemize}

Existing data does not further constrain the parameter space of the models
we discussed.
Current experiments, however, can test and distinguish these models.

\subsection{Predictions for $\boldsymbol{B\to K}$ data}

\begin{figure}[t]
\begin{center}
\includegraphics{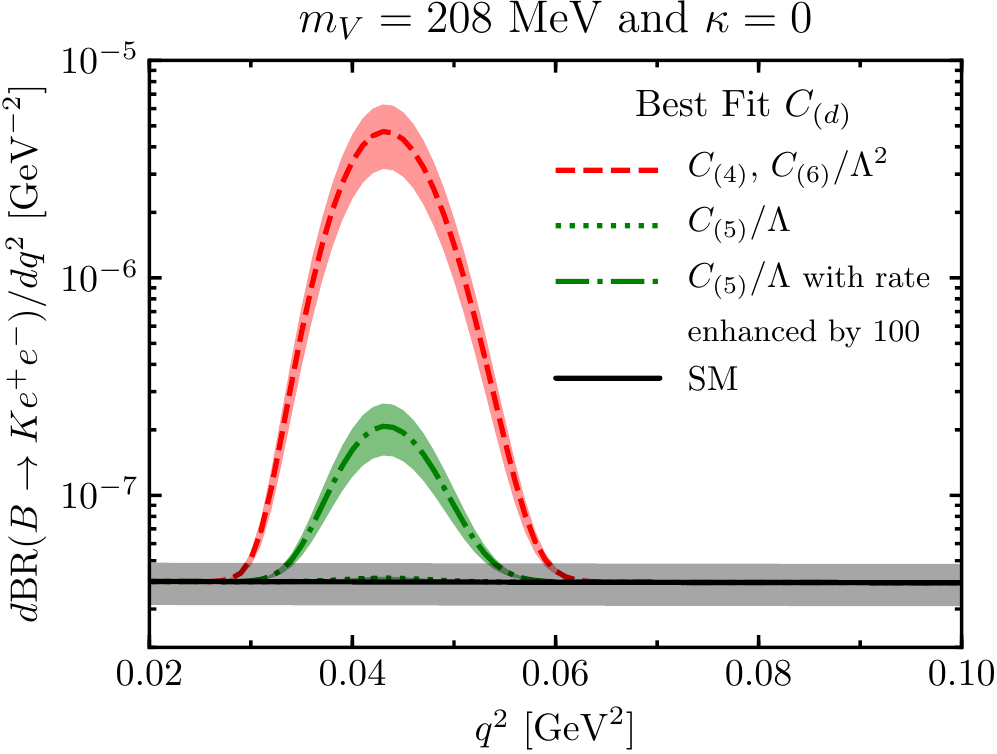}
\end{center}
\caption{The differential branching ratio  $d\text{BR}(B\to K e^+ e^-)/dq^2$
in the SM (solid black) and in the presence of a light vector resonance
with mass $208$ MeV produced via the operators $Q_{(4)}$ (dashed red),
$Q_{(5)}$ (dotted green) and
$Q_{(6)}$ (dashed red).
In each case, we use the best-fit value for the corresponding Wilson coefficients
(see \cref{tab:bestfitpoint}).
The predictions for $Q_{(4)}$ and $Q_{(6)}$ are identical.
The $Q_{(5)}$ peak is much less prominent.
For illustration we also show this case after enhancing the NP rate by a factor of
$100$ (dotted dashed green).
\label{fig:BRBKee}}
\end{figure}

Equipped with specific models we can now correlate the $B\to K^*$ results
of the previous subsection with currently possible $B\to K$ measurements.
We focus on the differential spectrum of $B \to K e^+e^-$ and $R_K$.
The relevant $B \to K V$ partial width induced by each of the operators
is
\begin{align}
\Gamma(B \to K V)\bigr\vert_{Q_{(4)}} &=
 	\frac{1}{64\pi} |C_{(4)}|^2 \frac{m_B^3}{m_V^2} \lambda^{3/2} f^2_{+}(m_V^2)
	\label{eq:BKrhoC4}~,\\
\Gamma(B \to K V) \bigr\vert_{Q_{(5)}} &=
	\frac{1}{16\pi} \frac{|C_{(5)}|^2}{\Lambda^2} m_B m_V^2 \left(1 + \frac{m_K}{m_B} \right)^{-2}  \lambda^{3/2} f_T^2(m_V^2) \label{eq:BKrhoC5}~,\\
\Gamma(B \to K V) \bigr\vert_{Q_{(6)}} &=
	\frac{1}{64\pi} \frac{|C_{(6)}|^2}{\Lambda^4} m_B^3 m_V^2 \lambda^{3/2} f^2_{+}(m_V^2)~,
	\label{eq:BKrhoC6}
\end{align}
where now $m_B$ denotes the $B^+$ mass and $m_{K}$ the $K^+$ mass.
In the  kinematical function $\lambda$ defined in \cref{eq:kinfactor}
$m_{K^*}$ should be replaced by $m_{K}$.
The $B \to K$ form factors $f_{+}$ and $f_{T}$ are taken
from ref.~\cite{Bailey:2015dka} (see \cref{sec:2bodyWidths}).

From eqs.~\eqref{eq:BKsrhoC4}--\eqref{eq:BKsrhoC6} and
eqs.~\eqref{eq:BKrhoC4}--\eqref{eq:BKrhoC6} we see that:
{(i)} The $B \to K V$ and $B \to K^* V$ decay widths
induced by $Q_{(5)}$ depend on different form factors than those induced
by $Q_{(4)}$/$Q_{(6)}$.
{(ii)} The $B \to K V$ and $B \to K^* V$ decay widths
induced by $Q_{(5)}$ have different scaling with the vector mass $m_V$, while those induced by $Q_{(4)}$/$Q_{(6)}$ do not.
Therefore, the magnitude of the peak in the $B\to K \ell^+\ell^-$ spectra
can be used as a way to disentangle the two production modes of the resonance.
{(iii)} The fact that
\begin{equation}
\frac{\Gamma(B\to K^*V)\bigr\vert_{Q_{(4)}}}{\Gamma(B\to K^*V)\bigr\vert_{Q_{(6)}}} =
\frac{\Gamma(B\to K  V)\bigr\vert_{Q_{(4)}}}{\Gamma(B\to K  V)\bigr\vert_{Q_{(6)}}}
\label{eq:C4C6relation}
\end{equation}
implies that the correlation of the $B\to K^*\ell^+\ell^-$ and $B\to K\ell^+\ell^-$
observables is identical in both production modes.
Therefore, the two production modes cannot be distinguished via the $B\to K \ell^+\ell^-$ spectra or a measurement of $R_K$ at
low $q^2$.

\begin{figure}[t]
\includegraphics{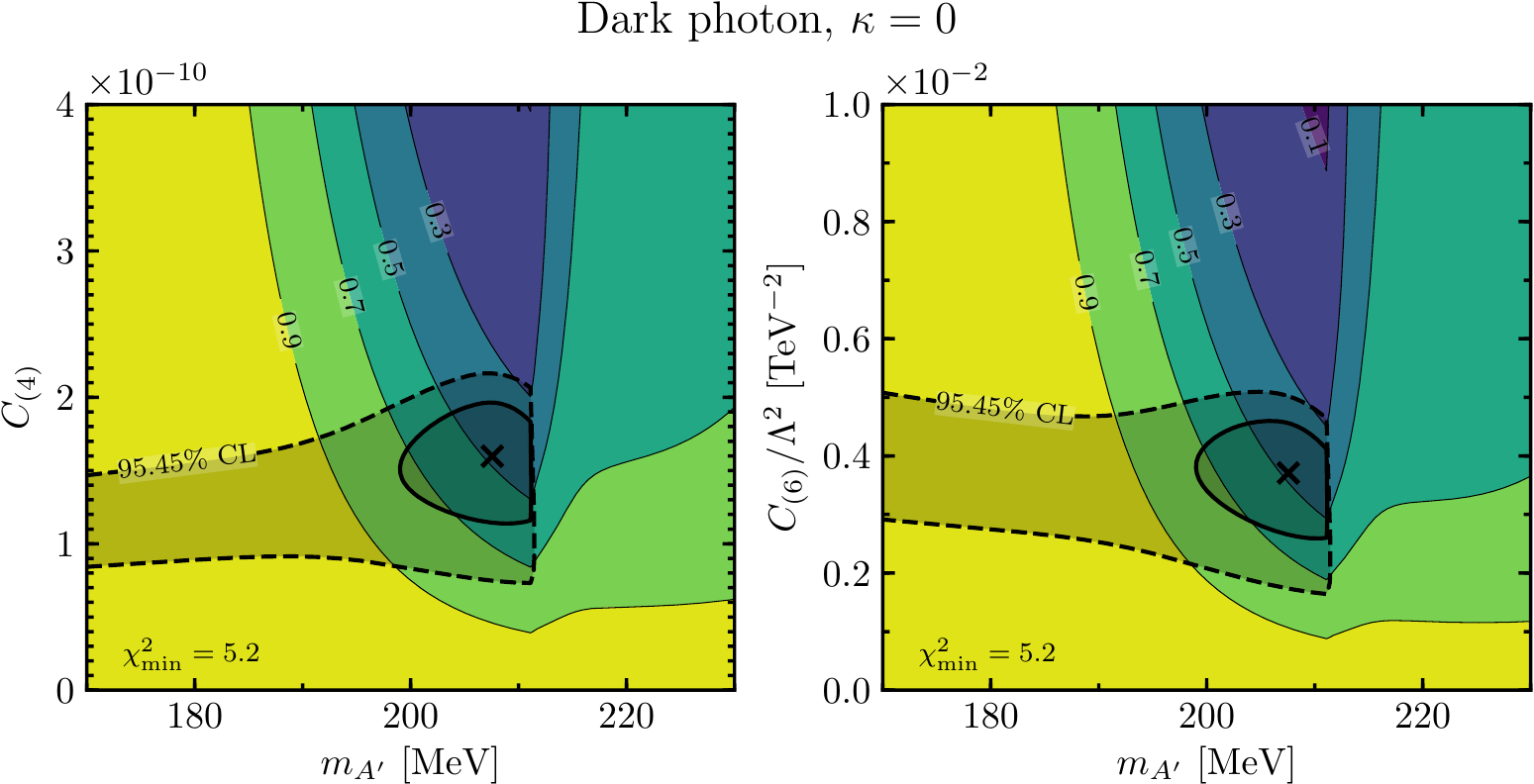}
\caption{
Values of $R_K$ for a dark photon with $\kappa=0$
as a function of the Wilson coefficients $C_{(4)}$ (left) and $C_{(6)}$ (right),
and the dark-photon mass, $m_{A'}$.
Superimposed is the best-fit region from the measurement of $R_{K*}$, $B\to K^{*} A' (\to \mu^+ \mu^-)$ and $B\to K^{*} e^+ e^-$.
We do not show the corresponding case for $Q_{(5)}$ because in the best-fit region
the effects in $R_K$ are unobservably small.
\label{fig:RKprediction}}
\end{figure}

Due to the absence of the photon-pole contribution to $B \to K e^+e^-$,
a peak in the $B \to K e^+e^-$ spectrum from the new resonance is
potentially even more prominent than in $B \to K^* e^+e^-$.
In \cref{fig:BRBKee} we show the differential
BR$(B \to K e^+e^-)$ as a function of $q^2$.
The solid black line depicts the predicted branching ratio in the SM,
computed using {\tt flavio}.
The red and green lines show the SM plus NP contribution from $Q_{(4)}/Q_{(6)}$ and
$Q_{(5)}$, respectively, at the best-fit points given in \cref{tab:bestfitpoint}.
The bands correspond to the $68.27$\,C.L.\ regions of BR$(B\to K^*V)$ from the
$\chi^2$ for the case $m_V=208$\,MeV.
The prediction from $Q_{(4)}$ and $Q_{(6)}$ coincide due to \cref{eq:C4C6relation},
which is why they are represented by the same line.
While $Q_{(4)}$ and $Q_{(6)}$ yield a sizeable deviation from the SM,
the contribution of $Q_{(5)}$ at the best-fit point
(dotted green line) is small compared to the SM theory uncertainties
(grey band).
The dotted dashed green line shows the $Q_{(5)}$ contribution
if the NP rate is enhanced by a factor $100$ with respect to the best-fit
rate.
The reason for this large suppression of the dipole contribution is
that the $B$ decay into the pseudoscalar $K$ and the vector $V$
via $Q_{(5)}$ is suppressed by $m_V^2/m_B^2$ compared to the decay
into the two vectors $K^*$ and $V$, due to angular-momentum conservation.

\begin{figure}[t]
\includegraphics{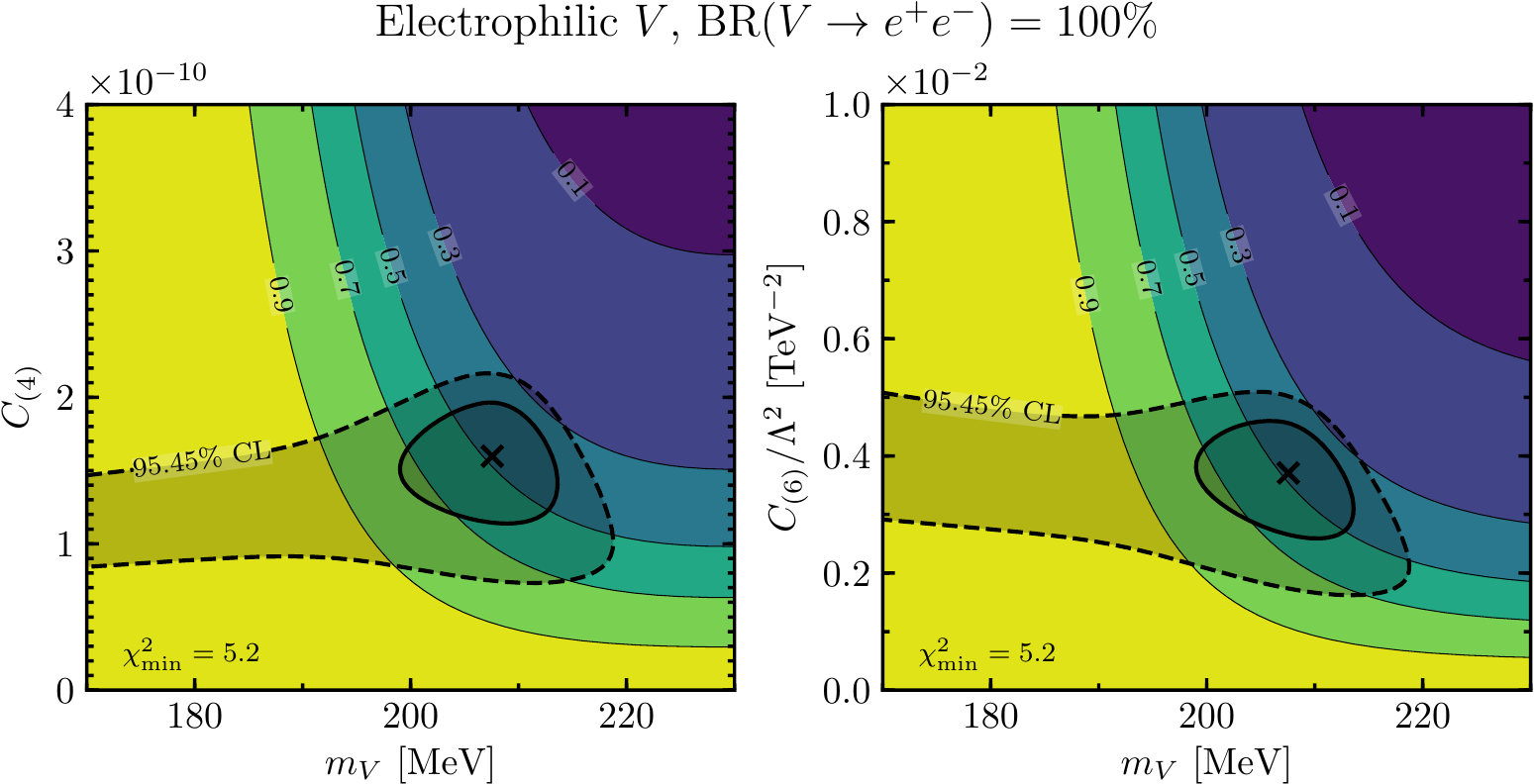}
\caption{Values of $R_K$ for an electrophilic $V$, i.e., BR$(V\to e^+ e^-)=100\%$,
as a function of the Wilson coefficients $C_{(4)}$ (left) and $C_{(6)}$ (right),
and the resonance mass, $m_{V}$.
Superimposed is the best-fit region from the measurement of $R_{K*}$
and the electron distribution.
We do not show the corresponding case for $Q_{(5)}$ because in the best-fit region
the effects in $R_K$ are unobservably  small.
\label{fig:electrophilic}}
\end{figure}

In \cref{fig:RKprediction}, we consider the case of a dark photon with $\kappa=0$.
We show contours of the predicted value of $R_K$ in a bin of $q^2\in[0.045,\,1]$\,GeV$^2$,
along with the $68.27\%$ and $95.45$\% C.L.\ regions
of the $\chi^2$ including the constraints from $B\to K^{*}A' (\to \mu^+\mu^-)$,
$B\to K^{*}e^+e^-$ and the measured value of $R_{K^*}$ in the low-$q^2$ bin.
We find that if the new resonance is produced via the operators $Q_{(4)}$ (left panel)
or $Q_{(6)}$ (right panel) then $R_K$ can be as low as $\sim 0.3$ in the $95.45$\%
C.L.\ preferred region.
If the new resonance is instead coupled via $Q_{(5)}$ then $R_K$ is barely altered
from its SM value and we do not show this case.

In \cref{fig:electrophilic}, we consider the case of an ``electrophilic''
vector resonance, i.e., BR$(V\to e^+e^-)=100\%$, and, analogously to \cref{fig:RKprediction},
show $R_K$ contours and the preferred regions from the $\chi^2$.
In this case, the bounds from the $B\to K^{*} V (\to \mu^+\mu^-)$ resonance search do not
apply and even a resonance with a mass above the di-muon threshold can account
for the low-$q^2$ bin of $R_K^*$ and significantly affect $R_K$.

For both the dark photon and the electrophilic case we see that,
if a future measurement of $R_K$ in such a low-$q^2$ bin finds
a value significantly smaller than the SM expectation,
the $Q_{(4)}$ and $Q_{(6)}$ production modes would be favored,
while the $Q_{(5)}$ mode would be disfavored.


\section{Cross-Checking Lepton-Universality Violation\label{sec:newLFU}}
The central issue looming over the subject of lepton universality
in semileptonic $B$ decays (and over NP speculations about
its origin) is the question of experimental uncertainties and of possible
unaccounted bias in the reconstruction of $e^+e^-$ pairs.
If the deviation in the low-$q^2$ bin of $R_{K^*}$ persists in the future, and at the same time no
peaks in the $B \to K^{(*)} e^+e^-$ and $B_s \to \phi e^+e^-$ spectra
are observed, the case for a systematic experimental origin of the deviation
would be strengthened.

The LHCb collaboration has performed detailed analyses of the
leptonic decays of $J/\psi$.
Those are known to be universal: ${\rm BR}(J/\psi\to \ell^+\ell^-)$
are equal for muon and electron final states to very
good accuracy~\cite{Ablikim:2013pqa}.
Therefore, LHCb uses these resonant sources of $\ell^+ \ell^-$
as a normalization for the continuum contribution in $R_K$ and $R_{K^*}$.
The collaboration also tests the overall consistency of the $e^+e^-$
reconstruction using photon conversion to electrons in the $K^*\gamma$
final states of $B^0$ decays.

Here we would like to point out that additional tests can and should
be made in other channels where one would not expect large deviations
from lepton universality, namely in decays to hadronic final states
with the lowest $\phi$ resonance, $m_\phi = 1020$\,MeV.
The $q^2$ value corresponding to $\phi \to \ell^+ \ell^-$ is $1.04$\,GeV$^2$
and is, therefore, very close to the interesting values for $q^2$.
$\phi$ mesons are copiously produced in a hadronic environment
and can be clearly seen as a peak in the di-muon invariant mass spectrum~\cite{LHCb:2016amn}.
However, in order to have the maximum resemblance to the semileptonic
$B$ decays, one should explore the
decay channels of charmed mesons that lead to charged hadrons and
a $\phi$, with $\phi$ decaying leptonically (e.g. $D^+\to\pi^\pm\mu^\mp\mu^+$~\cite{Aaij:2013sua}).

\begin{table}
\renewcommand{\arraystretch}{1.5}
\centering
\begin{tabular}{ l  c  c  c }
\hline\hline
\bf Decay mode 	&
\bf BR &
\bf Semileptonic BR, $\boldsymbol{\mu^+\mu^-}$ or $\boldsymbol{e^+e^-}$&
\bf $\boldsymbol{N_{\rm decays}}$ at 5\,fb$^{-1}$
\vspace{0.1em}\\
  \hline
$D^\pm \to \pi^\pm \phi$ & $5.4 \cdot 10^{-3}$ & $1.6\cdot 10^{-6}$ & ${\cal O}(10^4)$\\
$D^0\to \pi^+\pi^- \phi$ &  $2.6 \cdot 10^{-3}$ & $7.6 \cdot 10^{-7}$& ${\cal O}(10^4)$ \\
  $D_s^\pm \to \pi^\pm\phi $&  $2.5\cdot 10^{-2}$ & $1.3\cdot 10^{-5}$ & ${\cal O}(10^4)$ \\
  $D_s^\pm \to K^\pm\phi $ &  $1.8 \cdot 10^{-4}$ & $5.3\cdot 10^{-8}$ & ${\cal O}(10^2)$\\
   \hline\hline
\end{tabular}
\caption{Collection of $D^\pm,~D^0,$ and $D_s$ meson decay modes with which
tests of lepton universality of the $\phi$ meson
are possible at LHCb.
The individual branching ratios are extracted using PDG tables \cite{Patrignani:2016xqp},
while the leptonic branching to individual flavors is obtained by multiplying
with $BR(\phi\to \ell^+ \ell^-)$, which we take to be $2.9 \cdot 10^{-4}$.
The estimates for the number of expected events with $5$\,fb$^{-1}$ is obtained
by a simple rescaling of results from ref.~\cite{Aaij:2013sua}.
\label{tab:PhiDecayModes}}
\end{table}

In \cref{tab:PhiDecayModes}, we summarize the relevant decay
modes of charmed mesons that can be investigated by the LHCb collaboration.
Table \ref{tab:PhiDecayModes} suggests that the studies of leptonic
decays of $\phi$ generated by charmed mesons are entirely feasible given
the number of expected events. We take a previous study of
$D^\pm \to \pi^\pm \mu^+ \mu^-$ by LHCb, which recorded several
thousand $\phi$-mediated lepton pairs with $1$\,fb$^{-1}$ of integrated luminosity,
as an example~\cite{Aaij:2013sua} and make a simple rescaling to
higher integrated luminosity to estimate the number of expected events
with  $5$\,fb$^{-1}$.

Unlike the case of $B$ decays where continuum contributions are
comparable to the resonant one, the hadron $+$ $\ell^+ \ell^-$ decay
modes of $D$ mesons are dominated by resonances~\cite{Aaij:2013sua}.
Therefore, {\em if} the suggested test would produce highly discrepant
yields for lepton pairs from $\phi$ decays, e.g., by $\sim30\%$ as is
the case for $R_K$ and $R_{K^*}$,
then this would likely indicate a potential problem with the LHCb
reconstruction of electron pairs.
If on the other hand, the results for the $\phi$-mediated $\ell^+ \ell^-$
effects come out to be flavor universal, then it would further
strengthen the case for NP in $R_K$ and $R_{K^*}$.

As a note of potential curiosity, the comparison of the currently
most precise results for the leptonic widths of $\phi$ from
KLOE~\cite{Ambrosino:2004vg} and Novosibirsk~\cite{Akhmetshin:2010dt}
already produces a mildly non-universal answer at an approximately $10\%$ level.
In particular, taking the combination of
$\sqrt{\Gamma_{\phi\to ee}\times \Gamma_{\phi\to\mu\mu}}$ measured by
KLOE and combining it with the measurement of $\Gamma_{\phi\to ee}$
from Novosibirsk, we find ${\rm BR}(\phi\to \mu^+\mu^-)/ {\rm BR}(\phi\to e^+e^- ) = 1.15\pm  0.06$.
When we include all KLOE and Novosibirsk measurements of leptonic widths of $\phi$
the discrepancy is milder, ${\rm BR}(\phi\to \mu^+\mu^-)/ {\rm BR}(\phi\to e^+e^- ) =1.09 \pm 0.05$.


\section{Discussion and Conclusions\label{sec:conclusions}}
We explored possible NP explanations of the anomaly in the
low-$q^2$ bin of $R_{K^*}$ observed by LHCb. Heavy NP
parameterized in terms of an effective Lagrangian typically
does not affect the low-$q^2$ bin appreciably.
We found that effects from new, light degrees of freedom can
account for the observation, but are strongly constrained and
an explanation of the excess is only possible in a very narrow
range of parameter space.
In particular, we argued that off-shell exchange of a light resonance, $X$,
(significantly below the di-muon threshold)
can be excluded as the origin of the discrepancy at low $q^2$,
as the implied two-body decay rate $B \to X_s X$ typically
exceeds the measured total $B$ width.

An explanation in terms of one new resonance is possible if
the resonance mass is close to the di-muon threshold,
$m_X\simeq 2 m_\mu \simeq 211$~MeV, and if the resonance
decays predominantly into electrons.
Notably, in this mass range the difference of muon and
electron mass are enough to trigger effects in $R_{K^*}$
originating solely from kinematics,
without requiring a lepton-flavor non-universal coupling.
An simple interesting example model is given by a dark photon.
To explain the low-$q^2$ discrepancy one needs
BR$(B \to K^* A') \times$ BR$(A' \to e^+e^-) \sim 10^{-7}$.
The light resonance near the di-muon threshold affects mainly the low-$q^2$ bin of
$R_{K^*}$, while its effect at higher $q^2$ is negligible.
Additional NP is required to explain the high-$q^2$
bin of $R_{K^*}$ and the anomaly in $R_K$.

A fairly robust model-independent implication of a light-NP origin
of the low-$q^2$ discrepancy is a prominent peak close to the di-muon
threshold in the $B \to K^* e^+ e^-$ and $B_s \to \phi e^+ e^-$
di-electron invariant-mass spectra.
Within specific models, we investigated all possible couplings of a vector resonance up to
dimension six. We found that couplings from dimensions-four and five operators
can originate from plausible UV completions in the sense that their Wilson
coefficients may be induced from perturbative NP at or above the electroweak scale.
One possibility for a model for dimension-four couplings
is gauged (anomalous) baryon number with a gauge coupling of order $10^{-3}$.
This model does not require any sources of flavor violation beyond the SM,
neither in the quark nor the lepton sector.
It is also notable that the scale at which the dimension-five couplings are
induced in order to account for the low-$q^2$ $R_{K^*}$ anomaly
is compatible with the scale needed to explain its high-$q^2$ counterpart.

We also investigated the implications of these models for $B\to K$ data
and found that the size of a corresponding peak in the $B \to K e^+ e^-$
di-electron invariant-mass spectrum depends on the nature of the flavor violating
$b\to s$ coupling of the resonance. In particular, dimension-four and six
interactions lead to a prominent peak in $B \to K e^+ e^-$, while the dimension-five
interaction (dipole) leaves the $B \to K e^+ e^-$ decay SM-like to an
excellent approximation.

If the predicted peaks are not observed in future measurements, then this would
suggest that the effect is unlikely to originate from NP.
In such a case, a persistent anomaly in the low-$q^2$ bin
could imply a systematic experimental origin of the deviation,
which may also affect the interpretation of other
LFU violation hints, such as the high-$q^2$ bins of $R_{K^{(*)}}$.
Possible exotic NP explanations that would not predict a peak may still be
possible. These include unparticles, or a large discrete set of resonances that
are so close in mass that they cannot be resolved as peaks experimentally.

As an additional experimental cross check of LFU violation,
we proposed measurements of the leptonic $\phi$ branching ratios at LHCb.
To have the maximum resemblance to the semileptonic $B$ decays, we suggested to
explore the decay channels of charmed mesons to charged hadrons and a $\phi$.
We identified several $D^\pm,~D^0$, and $D_s$ meson decay modes, each
of which lead to ${\cal O}(10^4)$ leptonically decaying $\phi$'s with
$5$\,fb$^{-1}$ of data. This suggests excellent prospects for a
precise measurement of the ratio of $\phi \to \mu^+\mu^-$ and $\phi \to e^+e^-$
branching ratios.


\section*{Acknowledgements}
We thank Kaladi Babu and Pedro Machado for discussions.
WA and SG thank the Mainz Institute for Theoretical Physics (MITP) for its
hospitality and support during parts of this work.
The work of WA and SG was in part performed at the Aspen Center for Physics,
which is supported by National Science Foundation grant PHY-1607611.
The research of WA is supported by the National Science Foundation under
Grant No. PHY-1720252.
SG is supported by a National Science Foundation CAREER Grant No. PHY-1654502.
MJB and AT would like to thank Fermilab for its kind hospitality and support during the
early stages of this project.
Fermilab is operated by Fermi Research Alliance, LLC under Contract No.
DE-AC02-07CH11359 with the United States Department of Energy.
MJB was supported by the German Research Foundation (DFG) under Grant Nos. KO 4820/1-1 and
FOR 2239, by the European Research Council (ERC) under the European Union's Horizon 2020
research and innovation programme (grant agreement No. 637506, ``$\nu$Directions''), by
Horizon 2020 INVISIBLESPlus (H2020-MSCA-RISE-2015-690575) and by the Swiss National
Science Foundation (SNF) under contract 200021-175940.
The work of AT was supported under the International Cooperative Research and Development
Agreement for Basic Science Cooperation (CRADA No.~FRA-2016-0040) between Fermilab and
Johannes Gutenberg University Mainz, and partially by the Advanced Grant EFT4LHC of the
European Research Council (ERC) and the Cluster of Excellence Precision Physics,
Fundamental Interactions and Structure of Matter (PRISMA -- EXC 1098).

\appendix


\section{Light Off-Shell \texorpdfstring{$\boldsymbol{V}$}{V} in \texorpdfstring{$\boldsymbol{b \to s \ell \ell}$}{b -> s l l}\label{sec:3body}}

In this appendix we demonstrate that the measured
value of $R_{K^*}$ in the low-$q^2$ bin cannot be explained by
the off-shell exchange of a light vector boson, $V$, with vectorial
couplings to leptons and a mass significantly below the di-muon
threshold. Such a light vector could in principle lead to a
NP effect in the three-body decay $B \to K^* \ell^+ \ell^-$
that is enhanced at low $q^2$ by $m_B^2/q^2$.
In practice, however, we find that such an effect is severely
constrained by limits on the partial width of the inclusive
two-body decay $B \to X_s V$ and limits on the couplings of $V$
to leptons.

A robust limit on the $B \to X_s V$ partial width, which is
completely independent of the possible $V$ decay modes, is
given by the measured total $B$ width, $\Gamma(B \to X_s V) < 1/\tau_B$.
An equally robust and slightly stronger constraint can be obtained
from measurements of the charm yield per $B$ meson decay.
The BaBar analysis~\cite{Aubert:2006mp} finds that the average
number of charm quarks per $B^-$ decay is $N_c^- = 0.968^{+0.045}_{-0.043}$,
where we added the statistical uncertainty, the systematic uncertainty,
and the uncertainty from charm branching ratios in quadrature.
The measured value of $N_c^-$ implies that the branching ratios
of non-standard charmless decay modes such as $B \to X_s V$ are
bounded by $11.8\%$ at the $2\sigma$ level. It follows that
\begin{equation} \label{eq:2body_bound}
 \Gamma(B \to X_s V) \lesssim 11.8\% \times 1/\tau_B \simeq 4.7 \cdot 10^{-14}~\text{GeV} ~,
\end{equation}
where we used the lifetime of the charged $B^\pm$ meson
$\tau_B = 1.638 \pm 0.004$\,ps~\cite{Patrignani:2016xqp}.
We note that in many cases {\it much} stronger bounds on the
$B \to X_s V$ branching ratio can be obtained
depending on the $V$ decay modes.
If the $V$ decays dominantly to invisible final states or is stable
on detector scales, constraints from $B \to K^{(*)} \bar\nu\nu$ imply
BR$(B \to K V) \lesssim 1.7 \cdot 10^{-5}$~\cite{Lees:2013kla}
and BR$(B \to K^* V) \lesssim 4.0 \cdot 10^{-5}$~\cite{Lutz:2013ftz} at $90\%$
C.L.
Constraints at a similar level can be derived from
$B \to K^{(*)} e^+e^-$ measurements~\cite{Aaij:2013hha,Aaij:2015dea},
if the $V$ decays promptly into electrons and has a mass
$m_V \gtrsim 20$~MeV. We will not consider these much stronger
constraints in the following, as the model-independent constraint
in \cref{eq:2body_bound} turns out to be sufficiently
powerful to exclude observable effects in the low-$q^2$ bin of $R_{K^*}$.

In \cref{eq:Leff} we introduced the possible flavor violating
interactions of a new vector to SM quarks up to dimension six.
The contribution of the dimension-six interaction, $Q_{(6)}^{(\prime)}$,
to $B \to K^* \ell^+ \ell^-$ are not enhanced at low $q^2$ by $m_B^2/q^2$,
and we, therefore, only consider $Q_{(4)}$ and $Q_{(5)}$ in the following.
In the limit $m_{V} \ll m_B$, we find the following partial
decay widths of the inclusive decay $B \to X_s V$
\begin{align}
  \Gamma(B \to X_s V)\bigr\vert_{Q_{(4)}} = \frac{|C_{(4)}|^2}{32\pi} \frac{m_b^3}{m_{V}^2}\,,&
 &\Gamma(B \to X_s V)\bigr\vert_{Q_{(5)}} = \frac{|C_{(5)}|^2}{4\pi} \frac{m_b^3}{\Lambda^2}\,.
\end{align}
Using \cref{eq:2body_bound} and the PDG value for the bottom pole
mass $m_b = 4.78\pm 0.06$~GeV~\cite{Patrignani:2016xqp}, we find
the following bounds on the couplings $C_{(4)}$ and $C_{(5)}$
\begin{align}
 \label{eq:C4C5bound}
 |C_{(4)}| \lesssim \left(\frac{m_{V}}{100~\text{MeV}}\right) \times 2.1 \cdot 10^{-8}\,,&
 &\frac{|C_{(5)}|}{\Lambda} \lesssim 7.4 \cdot 10^{-8} ~\text{GeV}^{-1}\,.
\end{align}

\begin{figure}[t]
\includegraphics{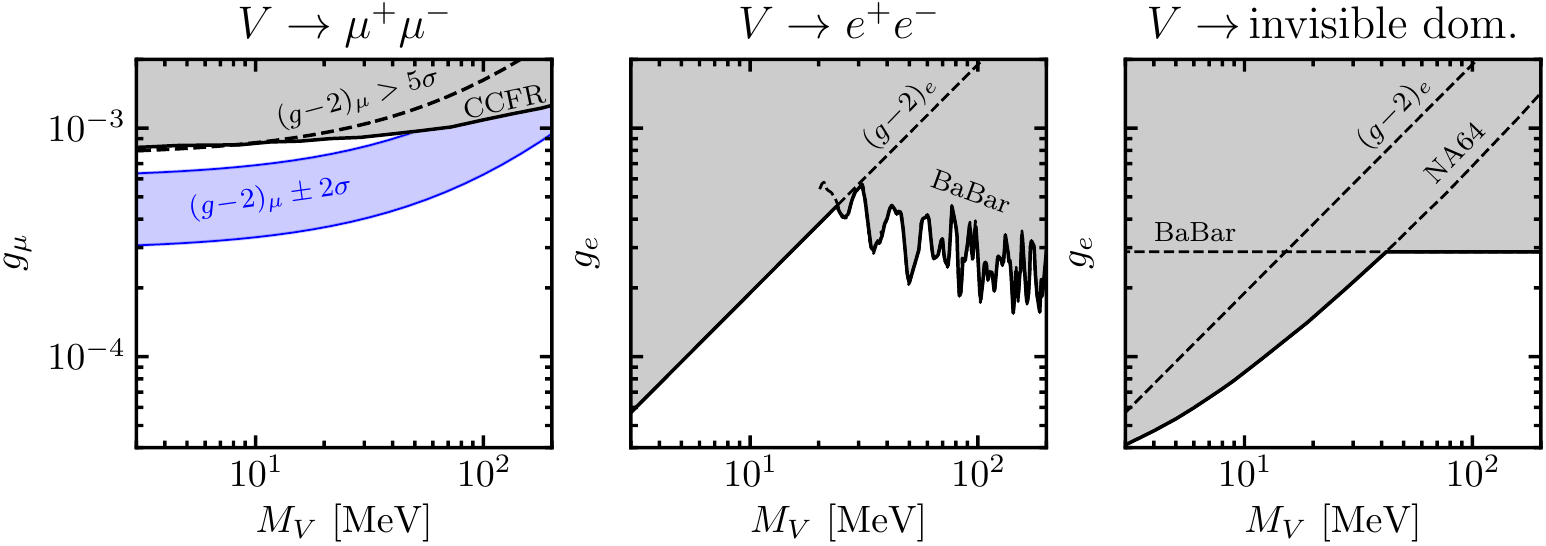}
\caption{Bounds on the vector couplings of a light vector,
$V$, to muons (left) and electrons (center and right) as a function of the vector mass.
On the left panel, the constraints from $\nu$-trident production based on
measurements by the CCFR collaboration
do not apply in models in which $V$ does not couple to neutrinos, i.e.,
dark-photon models.
The central plot assumes the absence of a relevant invisible decay
rate of the vector.
The right plot assumes that the invisible decays dominate.
\label{fig:lepton_couplings}}
\end{figure}

The off-shell corrections to the low-$q^2$ bin of $R_{K^*}$ do not
only depend on the quark flavor violating couplings of the $V$,
but also on the $V$ couplings to muons and electrons. Here we
focus on vector couplings\footnote{
Introducing simultaneously
axial-vector couplings and/or couplings from higher-dimensional
operators may open up the possibility of tuned cancellations
in some of the constraints discussed below. We do not consider
the possibility of such cancellations here.}
\begin{equation}
\mathcal{L}_\text{leptons} \supset g_e (\bar e \gamma^\nu e) V_\nu + g_\mu (\bar \mu \gamma^\nu \mu) V_\nu ~.
\end{equation}
Two concrete setups that induce such vector couplings are:
{(i)}  kinetic mixing of $V$ with the photon;
{(ii)} the gauging of flavor-specific lepton number.
In the case of kinetic mixing, one has $g_e = g_\mu$.
In the case of gauged lepton number, one simultaneously
also generates couplings to the corresponding neutrinos.
In the latter case, strong constraints on the coupling to muons,
$g_\mu$, can be derived from the measured rate of neutrino trident
production~\cite{Altmannshofer:2014pba}.
The bound is at the level of $g_\mu \lesssim  10^{-3}$ and is shown in
the left panel of \cref{fig:lepton_couplings}.
This bound is independent of the decay modes of the $V$.
Also shown in the plot is the region of parameter space that would
allow us to address the $(g-2)_\mu$ anomaly at the $2\sigma$ level,
as well as the exclusion by $(g-2)_\mu$ at the $5\sigma$ level.
The $(g-2)_\mu$ bound is independent of both the $V$ decay modes
and the couplings to neutrinos.

The anomalous magnetic moment of the electron, $(g-2)_e$,
leads to a bound on the $V$ coupling to electrons that is
independent of the $V$ decay modes.
The anomaly in the gyromagnetic
ratio of the electron, $a_e = \frac{1}{2}(g-2)_e$, can be predicted
in the SM with high precision using measurements of the fine-structure
constant in atomic physics experiments.
This results in the bound~\cite{Altmannshofer:2015qra}
\begin{equation}
 |\Delta a_e| \lesssim 8.1 \cdot 10^{-13} ~.
\end{equation}
The contribution to $a_e$ from a $V$ loop implies a bound on the coupling $g_e$
\begin{equation}
 \Delta a_e \simeq \frac{g_e^2}{12 \pi^2} \frac{m_e^2}{m_{V}^2} ~~ \Rightarrow ~~ g_e \lesssim \left( \frac{m_{V}}{100~\text{MeV}} \right)
 \times 1.9 \cdot 10^{-3}~,
\end{equation}
where we assumed $m_V \gg m_e$. Additional strong constraints on
the coupling to electrons can be obtained from the fixed target
experiment NA64~\cite{Banerjee:2016tad} and from BaBar
searches~\cite{Lees:2014xha,Lees:2017lec}. In the relevant range
of $V$ masses and couplings, there are only two possible decay
modes of $V$:
{(i)}  the coupling $g_e$ allows the $V$ to decay promptly to electrons;
{(ii)} $V$ can decay invisibly into neutrinos or a light dark sector.
If invisible decays are absent or negligibly small, the BaBar
search for dark photons~\cite{Lees:2014xha} leads to constraints
on $g_e$ that are stronger than the constraints from $(g-2)_e$ for
masses $m_{V} \gtrsim 20$~MeV.
In the central panel of \cref{fig:lepton_couplings} we show both constraints.
If the invisible decays dominate, the BaBar mono-photon search~\cite{Lees:2017lec}
and the NA64 search for dark photons~\cite{Banerjee:2016tad} lead
to strong constraints on $g_e$ as summarized in the right panel of
\cref{fig:lepton_couplings}.
Finally, if $V$ also has couplings to quarks (as in the dark-photon case)
additional constraints become relevant that restrict the parameter
space further~\cite{Essig:2013lka}.

\begin{figure}[t]
\includegraphics[]{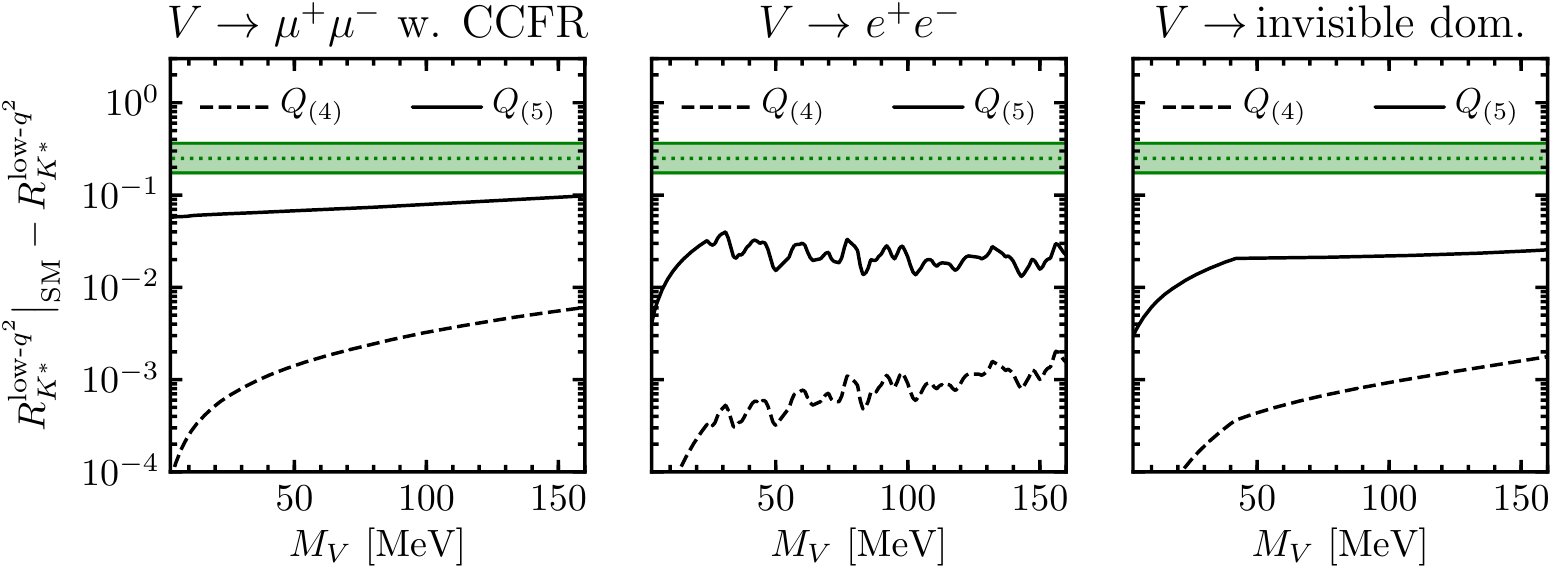}
\caption{
Maximal effects from $Q_{(4)}$ and $Q_{(5)}$
in the low-$q^2$ bin of $R_{K^*}$ from the off-shell
exchange of a light vector, $V$, as function of the vector mass.
In the left panel, $V$ decays solely to muons with a coupling
that saturates the $\nu$-trident bound from the left panel of
\cref{fig:lepton_couplings}.
In the central panel, $V$ decays solely to electrons
with a coupling saturating the bounds in the central
panel of \cref{fig:lepton_couplings}.
In the right panel, $V$ couples to electrons but primarily decays
invisibly, the maximally allowed coupling to electrons is plotted
in the right panel of \cref{fig:lepton_couplings}.
For each case, we show the maximal value of
$R_{K^*}^{[0.045,1.1]}\bigr\vert_\text{SM} - R_{K^*}^{[0.045,1.1]}\bigr\vert_{Q_{(d)}}$
as a funtion of the $V$ mass.
We see that the effects are much smaller than the current
discrepancy (horizontal green band).
\label{fig:max}}
\end{figure}

In \cref{fig:max} we show the maximal effects in the
low-$q^2$ bin of $R_{K^*}$ that can be induced by the off-shell
exchange of a light vector as function of the vector mass taking
into account the constraints on the couplings to quarks and leptons
discussed above.
We consider separately the production from $Q_{(4)}$ and $Q_{(5)}$,
taking their Wilson coefficients to saturate the bounds in \cref{eq:C4C5bound}.
To identify the maximal effect in $R_{K^*}$ we vary the sign of the Wilson
coefficients and the phase difference between the SM and the NP contribution.
In the left panel, we consider a muophilic case in which $V$ couples solely
to muon flavor.
For a given $V$ mass we take the maximally allowed $g_\mu$ coupling from the
$\nu$-trident bound (left panel of \cref{fig:lepton_couplings}).
In the central panel, we consider an electrophilic case in which $V$ decays solely
to electrons.
For a given $V$ mass we take the maximally allowed $g_e$ coupling from the
combination of the bounds in the central panel of \cref{fig:lepton_couplings}.
In the right panel, we consider the case in which $V$ can decay to electrons but
primarily decays invisibly.
For a given $V$ mass we take the maximally allowed $g_e$ coupling from the
combination of the bounds in the right panel of \cref{fig:lepton_couplings}.
For each case, we show the maximal value of
$R_{K^*}^{[0.045,1.1]}\bigr\vert_\text{SM} - R_{K^*}^{[0.045,1.1]}\bigr\vert_{Q_{(d)}}$
as a function of the $V$ mass.
We find that the effects are much smaller than the current
discrepancy
$R_{K^*}^{[0.045,1.1]}\bigr\vert_\text{SM} - R_{K^*}^{[0.045,1.1]}\bigr\vert_\text{LHCb} = 0.25 ^{+0.11}_{-0.08}$
(horizontal green band).


\section{Form Factors\label{sec:2bodyWidths}}
For the computation of the decay width $\Gamma(B \to K^{(*)} V)$
we use the form factors given in refs.~\cite{Straub:2015ica, Bailey:2015dka}.
In the limit of vanishing momentum transfer, $q^2 \rightarrow 0$,
the relevant $B \to K^*$ form factors are
\begin{align}
          V(0)   =\,& 0.341 \pm 0.036\,,&  A_1(0) =\,& 0.269 \pm 0.029\,,& &A_3(0) = 0.356 \pm 0.046\,,\nonumber\\[0.5em]
 T_1(0) = T_2(0) =\,& 0.282 \pm 0.031\,,&  T_3(0) =\,& 0.180 \pm 0.039\,,& &
\end{align}
and for the relevant $B \to K$ form factors
\begin{align}
f_+(0) = 0.335 \pm 0.036\,,&& f_T(0) = 0.279 \pm 0.067\,.
\end{align}
None of these form factors change appreciably
between $q^2=0$ and $q^2 = m_V^2 \sim 4m_\mu^2$.
In our numerics we thus use the zero-momentum transfer values above.

The functions $\mathcal{F}_{1}$ and $\mathcal{F}_{2}$ in
eqs.~\eqref{eq:BKsrhoC4}, \eqref{eq:BKsrhoC5}, and \eqref{eq:BKsrhoC6}
are given in terms of form factors by
\begin{equation} \label{eq:functionF1}
\begin{split}
\mathcal{F}_{1}\left(x,y\right)
 =&+ V^2\,     2 x y (1 - \sqrt{x})^2 \left(x^{2} - 2 x (1+y) + \left(1-y \right)^2 \right)\\
  &+ A_1^2\,   \frac{y}{4} (1 + \sqrt{x})^2 \left(-2 ( 3 x +1) y^{2} + (3 x + 1)^2 y + 8 (x - 1)^2 x + y^{3} \right) \\
  &+ A_3^2\,   x \left(x^{2} - 2 x (1+y) + \left(1-y \right)^2 \right)^2 \\
  &+ A_1 A_3\, (1 + \sqrt{x}) \sqrt{x} y (3 x - y + 1) \left(x^{2} - 2 x (1+y) + \left(1-y \right)^2 \right) \, ,
\end{split}
\end{equation}
\begin{equation} \label{eq:functionF2}
\begin{split}
\mathcal{F}_{2}\left(x,y\right)
=&+ T_1^2\,   8 x (1 - x)^2 \left(x^{2} - 2 x (1+y) + \left(1-y \right)^2 \right)\\
 &+ T_2^2\,   (1 - x)^2 \left(-2 ( 3 x +1) y^{2} + (3 x + 1)^2 y + 8 (x - 1)^2 x + y^{3} \right) \\
 &+ T_3^2\,   y \left(x^{2} - 2 x (1+y) + \left(1-y \right)^2 \right)^2 \\
 &+ T_2 T_3\, 2 (x - 1) y (3 x - y + 1) \left(x^{2} - 2 x (1+y) + \left(1-y \right)^2 \right) \, .
\end{split}
\end{equation}


\addcontentsline{toc}{section}{References}
\bibliographystyle{utphys}
\bibliography{references}

\end{document}